\documentclass[12pt,showpacs]{revtex4}
\usepackage[english]{babel}
\usepackage{graphicx}
\usepackage{epsfig}
\renewcommand{\d}{{\rm d}}

\def\be{\begin{equation}}
\def\ee{\end{equation}}
\def\bea{\begin{eqnarray}}
\def\eea{\end{eqnarray}}
\def\nn{\nonumber}
\def\eqcm{\: ,}                                          
\def\eqpt{\: .}
\def\adj{{\phantom{.}}}                           
\def\ket#1{\hbox{$\vert #1\rangle$}}   
\def\bra#1{\hbox{$\langle #1\vert$}}   
\def\oneh{{\textstyle {1\over 2}}}

\def          
\simeq{{\ \lower2pt\hbox{$-$}\mkern-13mu \raise2pt \hbox{$\sim$}\ }} 

\def\bold#1{\setbox0=\hbox{$#1$}%
      \kern-.02em\copy0\kern-\wd0
      \kern.04em\copy0\kern-\wd0
      \kern-.02em\raise.0433em\box0 }
\def\smbold#1{\setbox0=\hbox{$\scriptstyle#1$}%
      \kern-.02em\copy0\kern-\wd0
      \kern.04em\copy0\kern-\wd0
      \kern-.02em\raise.0433em\box0 }

\newcommand{\lcvec}[3]{\left[\;#1\;,\;#2\;,\;#3\;\right]}
\def\oggi{\number\day.\space 
\ifcase\month\or
 1.\or 2.\or 3.\or 4.\or 5.\or 6.\or
 7.\or 8.\or 9.\or 10.\or 11.\or 12.\fi
 \space\number\year}

\begin{document}
%
%
%


\title{Virtual meson cloud of the nucleon and generalized parton distributions
}
\author{B.~Pasquini and S.~Boffi
}
\affiliation{Dipartimento di Fisica Nucleare e Teorica, Universit\`a degli
Studi di Pavia and INFN, Sezione di Pavia, Pavia, Italy}
\date{\today}
\begin{abstract}
{We present the general formalism required to derive generalized parton
 distributions within a convolution model where the bare nucleon is dressed by its 
virtual meson cloud.  In the one-meson approximation the Fock states of the
 physical nucleon are expanded in a series  involving  a bare nucleon and 
two-particle, meson-baryon, states. 
The baryon is assumed here to be either a nucleon or a $\Delta$ 
described within the constituent quark model in terms of three valence quarks;
 correspondingly, the meson, assumed to be a pion, is described as a 
quark-antiquark pair. Explicit expressions for the unpolarized generalized
 parton distributions are obtained and evaluated in different kinematics.
}
\end{abstract}
\pacs{12.39.-x, 13.60.-r, 14.20.Dh}
\maketitle 
%


\section{Introduction}
\label{introduction}

The fundamental role of a nonperturbative pion cloud surrounding the nucleon is
 well explained in Quantum Chromodynamics (QCD) as a consequence of the 
spontaneously-broken chiral symmetry. The pion cloud associated with 
chiral-symmetry breaking was first discussed in the context of deep-inelastic 
scattering (DIS) by Feynman~\cite{feynman} and Sullivan~\cite{sullivan}. As 
realized by Thomas~\cite{thomas}, it can give an explanation of the 
flavor-symmetry 
violation in the sea-quark distributions of the nucleon thus naturally 
accounting for the excess of $\bar d$ (anti)quarks over $\bar u$ (anti)quarks 
as observed through the violation of the Gottfried sum 
rule~\cite{EMC,NA51,E866,Hermes}. Although the nucleon's nonperturbative 
antiquark sea cannot be ascribed entirely to its virtual meson 
cloud~\cite{Koepf}, the role of mesons in understanding these data 
 has been extensively discussed in connection with 
parton distributions (for reviews, see Refs.~\cite{Speth98,Londergan,Kumano}).

A similar important role is expected to be played by the meson cloud in the 
case of Generalized Parton Distributions (GPDs) that have recently been 
introduced and discussed in connection with Deeply Virtual Compton Scattering 
(DVCS) and hard exclusive meson production (for reviews, see 
Refs.~\cite{jig,radyushkin01,goeke,markusthesis}). Due to their intrinsic 
nonperturbative nature GPDs cannot be calculated directly within QCD. Up to 
now, only first results of their Mellin moments have been obtained by lattice
 calculations~(see, e.g., Ref.~\cite{lattice}). Therefore, in order to guide 
the planning of possible experiments one has still to rely on models.

The first calculation performed in the MIT bag model~\cite{bag} did not 
consider the pion cloud explicitly. Further calculations have been performed in
 the chiral quark-soliton model \cite{petrov,penttinen,schweitzer05,wakamatsu}. 
The model is based on
 an effective relativistic quantum field theory where the instanton 
fluctuations of the gluon field are simulated by a pion field binding the 
constituent quarks inside the nucleon. The model is theoretically justified in
 the limit of the large number of colors $N_c$, and in the leading order in 
the $1/N_c$ expansion it is not possible to obtain results for separate 
flavors, only special flavour combinations being nonzero. GPDs have also been
 calculated within the Nambu-Jona-Lasinio model~\cite{mineo}. A first attempt
 to explicitly include the meson cloud in a model calculation of GPDs has been 
discussed in Ref.~\cite{PBT} in terms of double 
distributions~\cite{radyushkin}.

A complete and exact overlap representation of GPDs has recently been worked 
out within the framework of light-cone quantization~\cite{DFJK,brodsky}. A 
Fock-state decomposition of the hadronic state is performed in terms of 
$N$-parton Fock states with coefficients representing the momentum light-cone
 wavefunction (LCWF) of the $N$ partons. The same approach has been followed 
in Refs.~\cite{BPT,BPT2,PPB}, 
where GPDs both in the chiral-even and chiral-odd 
sector were derived assuming that at the low-energy scale valence quarks can 
be interpreted as the constituent quarks treated in constituent quark models 
(CQMs). This assumption is based on the idea that there exists a scale $Q^2_0$ 
where the short range (perturbative) part of the interaction is negligible and,
 therefore, the glue and sea are suppressed, while a long range (confining) 
part of the interaction produces a proton composed by (three) valence quarks, 
mainly~\cite{PaPe}. Jaffe and Ross~\cite{JaRo} proposed to ascribe the quark
 model calculations of matrix elements to that hadronic scale $Q_0^2$.
In this 
way, quark models summarizing a great deal of hadronic properties may 
substitute for low-energy parametrizations, while evolution to larger momentum 
$Q^2$ is dictated by perturbative QCD. 

In this paper, we will study the possibility of integrating meson-cloud 
effects into the valence-quark contribution to GPDs.
Along the lines originally proposed in Refs.~\cite{Drell70,zoller}, 
a meson-baryon Fock-state expansion is used to construct 
the state $\ket{\tilde N}$ of the physical nucleon. 
In the one-meson approximation the state $\ket{\tilde N}$ is pictured as
 being part of the time a bare nucleon, $\ket{N}$, and part of the time a
 baryon-meson system, $\ket{BM}$. 
In this framework, it will be shown how to apply the convolution approach
used for the standard parton distributions in deep inelastic 
scattering~\cite{MSM}
to the case of GPDs. 
The main idea of the convolution approach is that there are no interactions
among the particles in a multi-particle Fock state during the interaction
with the hard photon. 
Therefore the external probe can scatter either on the bare nucleon, 
$\ket{N}$, or on one
of the constituents of the higher Fock states, $\ket{BM}$.
In the socalled DGLAP region, with one (anti)quark emitted and reabsorbed
by the physical nucleon,
the GPDs are obtained by folding
the quark GPDs within bare constituents 
(nucleons, pions, deltas, etc.) with the probability amplitudes describing the
distribution of these constituents 
in the dressed initial and final nucleon.
In contrast, in the socalled ERBL region where a $q\bar q$ pair is emitted 
from the initial nucleon, the GPDs are obtained from the overlap between
wavefunctions of Fock states with different parton number, which corresponds,
in the meson cloud model, to the
contribution of the interference terms between the $\ket{BM}$ component
in the initial state and the bare nucleon in the final state.

The model is revisited in Section~\ref{wf_mcm} where all the necessary 
ingredients that one can find scattered in the literature are consistently 
rederived to construct the LCWF in the meson cloud
 model.
The definition of the unpolarized GPDs are introduced in Section~\ref{gpds},
and the convolution formulas for the GPDs in the meson cloud model are 
explicitly derived in Section~\ref{gpds_mcm}.
In Section~\ref{model} we describe the model calculation considering the case 
of a pure pion cloud, and assuming a constituent quark model to construct
the LCWFs of the bare nucleon and the constituents 
of the baryon-meson Fock state.
 Numerical results are discussed in 
Section~\ref{results} and some conclusions are drawn in the final Section. 
Derivation of auxiliary quantities is detailed in three appendices.


\section{The meson-cloud model for the nucleon}
\label{wf_mcm}

The basic assumption throughout this paper is that the physical nucleon 
$\tilde N$ is made of a bare nucleon $N$ dressed by the surrounding meson 
cloud so that the state of the physical nucleon is decomposed according to the 
meson-baryon Fock-state expansion as a superposition of a bare nucleon state 
and states containing virtual mesons associated with recoiling baryons. 
This state, with four-momentum 
$p_N^\mu=(p^-_N,p^+_N,{\bf p}_{N\perp})\equiv(p^-_N,\tilde p_N)$ and helicity $\lambda$, is an eigenstate of the light-cone Hamiltonian
\begin{equation}
H_{LC}= \sum_{B,M} \left[H_0^B(q) + H_0^M(q)+H_I(N,BM)\right],
\label{eq:1}
\end{equation}
i.e.
\begin{equation}
H_{LC} |\tilde p_N,\lambda; \tilde N\rangle
=\frac{{\bf p}_{N\perp}^2+M_N^2}{p_N^+}|\tilde p_N,\lambda; \tilde N\rangle.
\label{eq:2}
\end{equation}
Here
$H_0^B(q)$ stands for the effective-QCD Hamiltonian which 
governs the constituent-quark dynamics, and leads to the confinement of 
three quarks in a baryon state;
analogously, $H_0^M(q)$ describes the quark interaction in a meson state. 
Thus we assume that the three- and two-quark states with the quantum numbers 
of a baryon and a meson are the eigenstates of $H_0^B(q)$ and 
$H_0^M(q)$, e.g.
\begin{eqnarray}
H_{0}^B |\tilde p_B,\lambda; B\rangle
=\frac{{\bf p}_{B\perp}^2+M_B^2}{p_B^+}|\tilde p_B,\lambda; B\rangle,
\label{eq:3}\\
H_{0}^M |\tilde p_M,\lambda; M\rangle
=\frac{{\bf p}_{M\perp}^2+M_M^2}{p_M^+}|\tilde p_M,\lambda; M\rangle.
\label{eq:4}
\end{eqnarray}
In Eq.~(\ref{eq:1}), $H_I(N,BM)$ is the
nucleon-baryon-meson interaction,
and the sum is over all the possible baryon and meson configurations in which
the nucleon can virtually fluctuate.
Using perturbation theory, we can expand the nucleon wavefunction 
in terms of the eigenstates of the bare Hamiltonian 
$H_0\equiv H_0^B(q) + H_0^M(q)$,
i.e.
\begin{eqnarray}
   |\tilde p_N,\lambda; \tilde N\rangle
& =& \sqrt{Z}\left(|\tilde p_N,\lambda; N\rangle
+{\sum_{n_1}}'\frac{|n_1\rangle \langle n_1 |H_I|\tilde p_N,\lambda; N\rangle}
{E_N-E_{n_1}+i\epsilon}\right.
\nonumber\\
& &\left.+{\sum_{n_1,n_2}}'\frac{|n_2\rangle \langle n_2 |H_I|n_1\rangle
\langle n_1| H_I|\tilde p_N,\lambda; N \rangle}{(E_N-E_{n_2}+i\epsilon)
(E_N-E_{n_1}+i\epsilon)} 
+\cdots\,\right) ,
\label{eq:5}
\end{eqnarray}
where $\sum'$ indicates the summation over $B M$ intermediate states, and 
$Z$ is the wavefunction
renormalization constant.
In the one-meson approximation, we truncate the series expansion of 
Eq.~(\ref{eq:5}) to the first order in $H_I$, and as a result we obtain
\begin{eqnarray}
|\tilde p_N,\lambda;\tilde N\rangle
&=& \sqrt{Z}|\tilde p_N,\lambda; N\rangle +\sum_{B,M}\,
|\tilde p_N, \lambda;N(BM)\rangle
\nn\\
&=&
\sqrt{Z}|\tilde p_N,\lambda; N\rangle 
+ 
\sum_{B,M}
\int \frac{{\rm d}p_B^+{\rm d}^2{\bf p}_{B\perp}}{ 2(2\pi)^3p_B^+} 
\int \frac{{\rm d}p_M^+{\rm d}^2{\bf p}_{M\perp}}{2(2\pi)^3p_M^+}\,
\nonumber\\
& &{}\ \times
\sum_{\lambda',\lambda''}
\frac{\langle B(\tilde p_B,\lambda')M(\tilde p_M,\lambda'') 
| H_I|N(\tilde p_N,\lambda)\rangle}
{E_N-E_B-E_M}\, 
|\tilde p_B,\lambda';B\rangle\,
|\tilde p_M,\lambda'';M\rangle. 
\label{eq:6}
\end{eqnarray}
In Eq.~(\ref{eq:6}), the normalization factor $\sqrt{Z}$ only affects the 
bare core $\ket{N}$, and not the meson-baryon component. 
As discussed in details in Refs.~\cite{MTh,Koepf},   
this prescription is consistent with assuming that the nucleon-baryon-meson
coupling constant in 
$H_I$ is taken equal to the renormalized value $g_{NBM},$ related
to lowest order to the
bare coupling $g^0_{NBM}$ via $g_{NBM}=\sqrt{Z}g^0_{NBM}$.

Finally, the hadron states in Eq.~(\ref{eq:6})
are normalized as
\begin{equation}
\langle p'^+,{\bf p}'_\perp, \lambda ';H\vert p^+,{\bf p}_\perp\lambda;H\rangle=
2(2\pi)^3p^+\delta(p'^+-p^+)\delta^{(2)}({\bf p}'_\perp -{\bf p}_\perp)
\delta_{\lambda\lambda'}.
\label{eq:7}
\end{equation}


\subsection{The nucleon wavefunction}
\label{wf}

In this Section starting from Eq.~(\ref{eq:6}) 
we derive the explicit general expression of the nucleon wavefunction 
on the basis of bare-nucleon and baryon-meson Fock states.

We first evaluate the energy denominator in Eq.~(\ref{eq:6}) using the 
following expression for the energy of the particles in terms of light-front 
variables 
\begin{equation}
E=\frac{1}{2}\left(p^+ +\frac{{\bf p}_\perp^2+M^2}{p^+}\right).
\label{eq:8}
\end{equation}
If we write the momenta of the baryon, $\tilde p_B$,  and the meson, 
$\tilde p_M$, in terms of the intrinsic (nucleon rest-frame) variables, i.e.
\be
\begin{array}{ll}
p^+_B= y p_N^+,\qquad &p^+_M=(1-y)p_N^+,\\
{\bf p}_{B\perp}={\bf k}_\perp+y\,{\bf p}_{N\perp},
\quad &{\bf p}_{M\perp}=-{\bf k}_\perp+(1-y)\,{\bf p}_{N\perp},
\end{array}
\label{eq:9}
\ee
we have
\begin{eqnarray}
(E_N-E_B-E_M)&=&\frac{1}{2p^+_N}\left(M_N^2-\frac{M_B^2+{\bf k}^2_\perp}{y}
-\frac{M_M^2+{\bf k}^2_\perp}{1-y}\right)\nn\\
&\equiv&
\frac{1}{2p^+_N}\left(M_N^2-M^2_{BM}(y,{\bf k}_\perp)\right),
\label{eq:10}
\end{eqnarray}
where
\begin{equation}
M^2_{BM}(y,{\bf k}_\perp)\equiv\frac{M_B^2+{\bf k}^2_\perp}{y}
+\frac{M_M^2+{\bf k}^2_\perp}{1-y},
\label{eq:11}
\end{equation}
is the invariant mass of the baryon-meson fluctuation.

Furthermore, the transition amplitude 
$\langle B(\tilde p_B,\lambda')M(\tilde p_M,\lambda'') | H_I|N(\tilde p_N,\lambda)\rangle$
in Eq.~(\ref{eq:6}) can be rewritten  as
\begin{eqnarray}
& &\langle B(\tilde p_B,\lambda')M(\tilde p_M,\lambda'') | H_I
|N(\tilde p_N,\lambda)\rangle=\nn\\
&&=
 (2\pi)^3\delta(p_B^+ +p^+_M -p^+_N)\,
\delta^{(2)}({\bf p}_{B\perp}+{\bf p}_{M\perp} -{\bf p}_{N\perp})
V^{\lambda}_{\lambda',\lambda''}(N,BM),
\label{eq:12}
\end{eqnarray}
where the vertex function $V^{\lambda}_{\lambda',\lambda''}(N,BM)$ 
has the following 
general expression
\be
V^{\lambda}_{\lambda',\lambda''}(N,BM)=
\bar u_{N \alpha}(\tilde p_N,\lambda) v^{\alpha\beta \gamma}
\chi_\beta(\tilde p_M,\lambda'')\psi_\gamma(\tilde p_B,\lambda').
\label{eq:13}
\ee
Here $u_N$ is the nucleon spinor, $\chi$ and $\psi$ are the field operators
of the intermediate meson and baryon, respectively, and
$\alpha,$ $\beta,$ $\gamma$ are bi-spinor and/or vector indices depending
on the representation used for particles of given type.

Using the results of Eqs.~(\ref{eq:10}) and (\ref{eq:12}), 
we find
\begin{eqnarray}
|\tilde p_N,\lambda;\tilde N\rangle
& &
=\sqrt{Z}|\tilde p_N,\lambda; N\rangle
+
\sum_{B,M}
\int \frac{{\rm d}y{\rm d}^2{\bf k}_{\perp}}{2(2\pi)^3}\,
\frac{1}{\sqrt{y(1-y)}}
\sum_{\lambda',\lambda''}
\phi_{\lambda'\lambda''}^{\lambda \,(N,BM)}(y,{\bf k}_\perp)\nonumber\\
& &\quad{}\times
|yp^+_N,{\bf k}_{\perp}+y{\bf p}_{N\perp},\lambda';B\rangle\,
|(1-y)p^+_N,-{\bf k}_{\perp}+(1-y){\bf p}_{N\perp},\lambda'';M\rangle,
\label{eq:14}
\end{eqnarray}
where we introduced the function 
$\phi_{\lambda'\lambda''}^{\lambda\,(N,BM)}(y,{\bf k}_\perp)$ to define the 
probability amplitude for a nucleon with helicity $\lambda$ to fluctuate into 
a virtual $BM$ system with the baryon having helicity $\lambda'$, longitudinal 
momentum fraction $y$ and transverse momentum ${\bf k}_\perp$, and the meson 
having helicity $\lambda''$, longitudinal momentum fraction $1-y$ and 
transverse momentum $
-{\bf k}_\perp$, i.e.
\begin{eqnarray}
\phi_{\lambda'\lambda''}^{\lambda\,(N,BM)}(y,{\bf k}_\perp)
=\frac{1}{\sqrt{y(1-y)}}\,
\frac{V^\lambda_{\lambda',\lambda''}(N,BM)}
{M^2_N-M^2_{BM}(y,{\bf k}_\perp)}.\,
\label{eq:15}
\end{eqnarray}
We note that Eq.~(\ref{eq:14}) is equivalent to the expression of the nucleon 
wavefunction obtained in the framework of  ``old-fashioned'' time-ordered 
perturbation theory in the infinite momentum frame (see Ref.~\cite{Drell70}).  

By imposing the normalization of the nucleon state as in Eq.~(\ref{eq:7}), 
from Eq.~(\ref{eq:14}) we obtain the following condition on the normalization factor $Z$
\be
1=Z+P_{BM/N},
\label{eq:16}
\ee
with
\be
P_{BM/N}=\sum_{B, M}
\int \frac{{\rm d}y{\rm d}^2{\bf k}_{\perp}}{2(2\pi)^3}\,
\frac{1}{y(1-y)}
\sum_{\lambda',\lambda''}\frac{|V^{1/2}_{\lambda',\lambda''}(N,BM)|^2}
{[M^2_N-M^2_{BM}(y,{\bf k}_\perp)]^2}.
\label{eq:17}
\ee
Here
$P_{BM/N}$ is the probability of fluctuation of the nucleon in a
 baryon-meson state, and, accordingly, $Z$ gives  the probability to 
find the bare
nucleon in the physical nucleon.  


\subsection{Partonic content of the nucleon wavefunction}
\label{partons}

In this Section we derive the expression of the nucleon wavefunction 
(\ref{eq:14}) in terms of the constituent partons
of the nucleon core and of the meson-baryon components. 

The light-front state of the bare nucleon is given by
\be
\ket{{\tilde{p}}_N,\lambda;N} = \sum_{\tau_i,\lambda_i}
\int\left[\frac{{\rm d}x }{\sqrt{x}}\right]_3
[{\rm d}^2{\bf k}_\perp]_3
\Psi_\lambda^{N,[f]}(\{x_i,{\bf k}_{\perp i};\lambda_i,\tau_i\}_{i=1,2,3})
\prod_{i=1}^3
\ket{x_i p^+_N, \, {\bf p}_{i\perp},\lambda_i,\tau_i;q},
\label{eq:18}
\ee
where $\Psi_\lambda^{N,[f]}(\{x_i,{\bf k}_{\perp i};\lambda_i,\tau_i\}_{i=1,2,3})$ is the momentum light-cone wavefunction  which gives the probability 
amplitude for finding
in the nucleon three quarks with momenta
 $(x_i p^+_N, {\bf p}_{i\perp}=\,{\bf k}_{i\perp} + x_i {\bf p}_{N\perp})$, 
and spin and isospin variables $\lambda_i$ and $\tau_i,$ respectively. 
In Eq.~(\ref{eq:18}) and in the following formulas, 
the integration measures are defined by
\be
\label{eq:19}
\left[\frac{{\rm d} x}{\sqrt{x}}\right]_N 
= \left(\prod_{i=1}^N \frac{{\rm d} x_i}{\sqrt{x_i}}\right)
\delta\left(1-\sum_{i=1}^N x_i\right),
\ee
\be
\label{eq:20}
[{\rm d}^2{\bf k}_\perp]_N = \left(\prod_{i=1}^N
\frac{{\rm d}^2{\bf k}_{\perp\,i}}{2(2\pi)^3}\right)\,2(2\pi)^3\,
\delta\left(\sum_{i=1}^N {\bf k}_{\perp\,i}\right).
\ee

Next we consider the component of the meson-baryon Fock state 
in Eq.~(\ref{eq:14}).

The light-front state of the baryon is given by
\be
\ket{ \tilde{p}_B,\lambda';\, B} = \sum_{\tau_i,\lambda_i}
\int\left[\frac{{\rm d} x}{\sqrt{x}}\right]_3
[{\rm d}^2{\bf k}_\perp]_3
\Psi_{\lambda'}^{B,\, [f]}(\{x_i,{\bf k}_{\perp i};\lambda_i,\tau_i\}_{i=1,2,3})
\prod_{i=1}^3
\ket{x_i p^+_B, \, {\bf p}_{i\perp},\lambda_i,\tau_i;q},
\label{eq:21}
\ee
where now the intrinsic variables of the quarks $x_i$ and $k^+_i$ 
refer to the baryon rest frame, i.e. $x_i=p^+_i/p_B^+$ and
 ${\bf p}_{i\perp}= {\bf k}_{i\perp}+x_i{\bf p}_{B\perp}$ ($i=1,2,3$).

An analogous expression holds for the light-front state of the meson, i.e.
\begin{eqnarray}
\ket{\tilde{p}_M,\lambda'';\, M} &=& \sum_{\tau_i,\lambda_i}
\int
\frac{\d x_4 \d x_5}{\sqrt{x_4 x_5}}
\frac{\d{\bf k}_{4\perp}
\d{\bf k}_{5\perp}}{16\pi^3}\,\delta(1-x_4-x_5)
\,\delta^{(2)}({\bf k}_{4\perp}+{\bf k}_{5\perp})\nn\\
& &{}\ \times
\Psi^{M,\, [f]}_{\lambda''}(\{x_i,{\bf k}_{\perp i};\lambda_i,\tau_i\}_{i=4,5})
\prod_{i=4}^5
\ket{x_i p^+_M, \, {\bf p}_{i\perp},\lambda_i,\tau_i;q},
\label{eq:22}
\end{eqnarray}
with $x_i=p^+_i/p_M^+$, 
${\bf p}_{i\perp}= x_i{\bf p}_{M\perp}+{\bf k}_{i\perp}$ ($i=4,5$).

When we insert the expressions of the baryon and meson states in 
Eq.~(\ref{eq:14}), it is convenient to rewrite the kinematical variables of 
the partons as follows.

For $ i=1,2,3$:
\begin{eqnarray*}
x_i &= &\displaystyle \frac{p^+_i}{p^+_B}=\frac{p^+_i}{p^+_N}\frac{p^+_N}{p^+_B}=\frac{\xi_i}{y},\nn \\
{\bf p}_{i\perp}& =&x_i{\bf p}_{B\perp}+{\bf k}_{i\perp}
=x_i\,({\bf k}_\perp+y\,{\bf p}_{N\perp})+{\bf k}_{i\perp}\nn\\
&=&\xi_i{\bf p}_{N\perp}+{\bf k}_{i\perp}+x_i\,{\bf k}_\perp\equiv
\xi_i{\bf p}_{N\perp}+{\bf k}'_{i\perp},
\end{eqnarray*}
where $\xi_i=p^+_i/p_N^+$ is the fraction of the longitudinal momentum of the
 nucleon carried by the quarks in the baryon, and 
${\bf k}'_{i\perp}$ is the intrinsic 
transverse momentum of the quarks with respect to the nucleon rest frame.

For $i=4,5$:
\begin{eqnarray*}
x_i &=& \displaystyle \frac{p^+_i}{p^+_M}=\frac{p^+_i}{p^+_N}\frac{p^+_N}{p^+_M}=\frac{\xi_i}{1-y},\nn\\
{\bf p}_{i\perp}&=&x_i{\bf p}_{M\perp}+{\bf k}_{i\perp}
=x_i\,(-{\bf k}_\perp+(1-y)\,{\bf p}_{N\perp})+{\bf k}_{i\perp}\nn\\
&=&\xi_i{\bf p}_{N\perp}+{\bf k}_{i\perp}-x_i\,{\bf k}_\perp
\equiv\xi_i{\bf p}_{N\perp}+{\bf k}'_{i\perp},
\end{eqnarray*}
with $\xi_i=p^+_i/p_N^+$ and ${\bf k}'_{i\perp}$ the intrinsic variables
of the quarks in the meson   with respect to the nucleon rest frame.
Accordingly we transform the variables of integration as follows.

For $ i=1,2,3$:
\begin{eqnarray*}
x_i &\rightarrow& \xi_i=y\, x_i ,\nn\\
{\bf k}_{i\perp}&\rightarrow &{\bf k}'_{i\perp}={\bf k}_{i\perp}+x_i\,{\bf k}_\perp.
\end{eqnarray*}

For $i=4,5$:
\begin{eqnarray*}
x_i &\rightarrow& \xi_i=(1-y)\, x_i ,\nn\\
{\bf k}_{i\perp}&\rightarrow &{\bf k}'_{i\perp}={\bf k}_{i\perp}-x_i\,{\bf k}_\perp.
\end{eqnarray*}

The meson-baryon component of the nucleon wavefunction in Eq.~(\ref{eq:14}) can then be written as
\begin{eqnarray}
|\tilde p_N, \lambda;N(BM)\rangle
&=&
\int {\rm d}y\,{\rm d}^2{\bf k}_{\perp}\,
\int_0^y \prod_{i=1}^3\, \frac{\d\xi_i}{\sqrt{\xi_i}}\,
\int_0^{1-y} \prod_{i=4}^5\, \frac{\d\xi_i}{\sqrt{\xi_i}}\,
\int \prod_{i=1}^5\, \frac{\d{\bf k}'_{i\perp}}{[2(2\pi)^3]^4}\,\nn\\
& &{}\times
\delta\left(y-\sum_{i=1}^3\xi_i\right)
\delta^{(2)} \left({\bf k}_\perp-\sum_{i=1}^3{\bf k}'_{i\perp} \right)\,
\delta \left(1-\sum_{i=1}^5\xi_i \right)\,
\delta^{(2)} \left(\sum_{i=1}^5{\bf k}'_{i\perp} \right)\nn\\
& &{}\times
\sum_{\lambda',\lambda''}\sum_{\lambda_i,\tau_i}
\frac{V^\lambda_{\lambda',\lambda''}(N,BM)}
{M^2_N-M^2_{BM}(y,{\bf k}_\perp)}
\tilde \Psi_{\lambda'}^{B,\,[f]}(\{\xi_i,{\bf k}'_{i\perp};\lambda_i,\tau_i\}_{i=1,2,3})\nn\\
& &{}\times 
\tilde \Psi_{\lambda''}^{M,\,[f]}(\{\xi_i,{\bf k}'_{i\perp};\lambda_i,\tau_i\}_{i=4,5})
\prod_{i=1}^5\,
|\xi_ip^+_N,\,{\bf k}'_{\perp}+\xi_i{\bf p}_{N\perp},\,\lambda_i,\,
\tau_i;q\rangle,
\label{eq:23}
\end{eqnarray}
where the wavefunctions $\tilde \Psi_{\lambda'}^{B,\,[f]}$ and 
$\tilde \Psi_{\lambda''}^{M,\,[f]}$ incorporate the Jacobian  ${\cal J}$ of
 the transformation $x_i\rightarrow \xi_i$, i.e.
\begin{eqnarray}
\tilde \Psi_{\lambda'}^{B,\,[f]}(\{\xi_i,{\bf k}'_{i\perp};\lambda_i,\tau_i\}_{i=1,2,3})
&=&\sqrt{{\cal J}(\xi_1,\xi_2,\xi_3)}
\tilde\Psi_{\lambda'}^{B,\,[f]}(\{x_i,{\bf k}_{i\perp};\lambda_i,\tau_i\}_{i=1,2,3})
\nn\\
&=&
\frac{1}{y^{3/2}}
\tilde\Psi_{\lambda'}^{B,\,[f]}(\{x_i,{\bf k}_{i\perp};\lambda_i,\tau_i\}_{i=1,2,3}),
\label{eq:24}
\end{eqnarray}
\begin{eqnarray}
\tilde \Psi_{\lambda''}^{M,\,[f]}(\{\xi_i,{\bf k}'_{i\perp};\lambda_i,\tau_i\}_{i=4,5})
&=&\sqrt{{\cal J}(\xi_4,\xi_5)}
\tilde\Psi_{\lambda''}^{M,\,[f]}(\{x_i,{\bf k}_{i\perp};\lambda_i,\tau_i\}_{i=4,5})
\nn\\
&=&\frac{1}{(1-y)}
\tilde\Psi^{M,\,[f]}(\{x_i,{\bf k}_{i\perp};\lambda_i,\tau_i\}_{i=4,5}).
\label{eq:25}
\end{eqnarray}
Finally, by introducing the following definition 
\begin{eqnarray}
& &\tilde\Psi^{5q,[f]}_\lambda
(y,{\bf k}_\perp; \,\{\xi_i,{\bf k}'_{i\perp};\lambda_i,\tau_i\}_{i=1,...,5})
\equiv
\sum_{\lambda',\lambda''}
\frac{V^\lambda_{\lambda',\lambda''}(N,BM)}
{M^2_N-M^2_{BM}(y,{\bf k}_\perp)}\nn\\
& &\qquad\qquad\qquad\qquad{}\times 
\tilde \Psi_{\lambda'}^{B,\,[f]}(\{\xi_i,{\bf k}'_{i\perp},\lambda_i,\tau_i\}_{i=1,2,3})
\tilde \Psi_{\lambda''}^{M,\,[f]}(\{\xi_i,{\bf k}'_{i\perp},\lambda_i,\tau_i\}_{i=4,5}),
\label{eq:26}
\end{eqnarray}
Eq.~(\ref{eq:23}) can be simplified to the following expression:
\begin{eqnarray}
|\tilde p_N, \lambda;N(BM)\rangle
&=&
\int {\rm d}y\,{\rm d}^2{\bf k}_{\perp}\,
\int_0^y \prod_{i=1}^3\, \frac{\d\xi_i}{\sqrt{\xi_i}}\,
\int_0^{1-y} \prod_{i=4}^5\, \frac{\d\xi_i}{\sqrt{\xi_i}}\,
\int \prod_{i=1}^5\, \frac{\d{\bf k}'_{i\perp}}{[2(2\pi)^3]^4}\nn\\
& &{}\times
\delta\left(y-\sum_{i=1}^3\xi_i\right)
\delta^{(2)} \left({\bf k}_\perp-\sum_{i=1}^3{\bf k}'_{i\perp}\right)\,
\delta \left(1-\sum_{i=1}^5\xi_i \right)\,
\delta^{(2)} \left(\sum_{i=1}^5{\bf k}'_{i\perp} \right)\nn\\
& &{}\times 
\sum_{\lambda_i,\tau_i}
\tilde \Psi_{\lambda}^{5q,\,[f]}(y,{\bf k}_\perp; \,
\{\xi_i,{\bf k}'_{i\perp};\lambda_i,\tau_i\}_{i=1,...,5})\nn\\
& &{}\times 
\prod_{i=1}^5\,
|\xi_ip^+_N,\,{\bf k}'_{\perp}+\xi_i{\bf p}_{N\perp},\,\lambda_i,\,
\tau_i;q\rangle,
\label{eq:27}
\end{eqnarray}
where $\tilde \Psi_{\lambda}^{5q,\,[f]}$ can be interpreted as the probability 
amplitude for finding in the nucleon
a configuration of five partons composed by 
two clusters of three and two quarks, with total momentum
$(y p^+_N, {\bf p}_{B\perp})$ and    
$((1-y) p^+_N, {\bf p}_{M\perp}),$ respectively.


\section{The unpolarized generalized parton distributions}
\label{gpds}

In the definition of GPDs it is useful to choose a symmetric frame of 
reference where the virtual photon momentum $q^\mu$ and the average nucleon 
momentum $\bar p_N^\mu=\oneh(p_N^\mu+{p'}_N^\mu)$ are collinear along 
the $z$ axis and in opposite  directions, i.e.
\begin{eqnarray} 
p_N  &=& 
\lcvec{\frac{m^2+{\bf\Delta}_\perp^2/4}{2(1+\xi)\,\bar p_N^{\,+}}}
      {(1+\xi)\bar p_N^{\,+}}
      {-\frac{{\bf\Delta}_\perp}{2}} 
      \equiv \left[\frac{m^2+{\bf\Delta}_\perp^2/4}{2(1+\xi)\,\bar p_N^{\,+}}\,,{\tilde p}_N\right], \nn \\
p_N^{\prime} &=& 
\lcvec{\frac{m^2+{\bf\Delta}_\perp^2/4}{2(1-\xi)\,\bar p_N^{\,+}}}
      {(1-\xi)\,\bar p_N^{\,+}}
      {+\frac{{\bf\Delta}_\perp}{2}}
      \equiv\left[\frac{m^2+{\bf\Delta}_\perp^2/4}{2(1-\xi)\,\bar p_N^{\,+}}\,,{\tilde p}'_N\right].
\label{eq:28}
\end{eqnarray}
Furthermore, $Q^2 = -q^\mu q_\mu$ is the space-like virtuality that defines the scale of the process,  $t=\Delta^2=({p'}_N^\mu-p_N^\mu)^2$ is the invariant transferred momentum square, and the skewness $\xi$ describes the longitudinal change of the nucleon momentum,  $2\xi=-\Delta^+/\bar p^+_N$.

According to Ref.~\cite{jig}, for each flavor $q$ the soft amplitude corresponding to unpolarized GPDs reads 
\be
\label{eq:29}
F^q_{\lambda'_N\lambda_N}(\bar x,\xi,{\bf \Delta}_\perp) =
\left. \frac{1}{2\sqrt{1-\xi^2}}
\int \frac{{\rm d}z^-}{2\pi}\, e^{i\bar x\,\bar p_N^+y^-}
\bra{p'_N,\lambda'_N}\overline\psi(-\oneh z)\,\gamma^+\, 
\psi(\oneh z)\ket{p_N,\lambda_N}
\right\vert_{z^+={\bf z}_\perp=0},
\ee
where $\bar x$ defines the fraction of the quark light-cone momentum 
(${\overline k}^+={\bar x}\,{\bar p}_N^+$), $\lambda_N$
 ($\lambda'_N$) is the helicity of the initial (final) nucleon, and 
the quark-quark correlation function is integrated along the light-cone 
distance $z^-$ at equal light-cone time ($y^+=0$) and zero transverse 
separation (${\bf z}_\perp=0$) between the quarks. The leading twist 
(twist-two) part of this amplitude can be parametrized as 
\bea
F^q_{\lambda'_N\lambda_N}(\bar x,\xi,{\bf \Delta}_\perp) 
& = & \frac{1}{2\bar p_N^{\,+}\sqrt{1-\xi^2}}
\, \bar u(\tilde p'_N,\lambda'_N)\,\gamma^+\, u(\tilde p_N,\lambda_N) \, 
H^q(\bar x,\xi,{\bf \Delta}_\perp)
 \nonumber\\
& &{} + \frac{1}{2\bar p_N^{\,+}\sqrt{1-\xi^2}} 
\, \bar u(\tilde p'_N,\lambda'_N)\,
\frac{i\sigma^{+\nu}\Delta_\nu}{2M_N}  \,u(\tilde p_N,\lambda_N)
\,E^q(\bar x,\xi,{\bf \Delta}_\perp) ,
\\ \nonumber
\label{eq:30}
\eea
where $H^q(\bar
x,\xi,{\bf \Delta}_\perp)$ and  $E^q(\bar x,\xi,{\bf \Delta}_\perp)$ are
 the chiral-even helicity conserving and  helicity flipping GPDs for partons 
of flavor $q$, respectively. Taking different proton-helicity combinations, we
 have
\bea
F^q_{++}(\bar x,\xi,{\bf \Delta}_\perp) & = & 
F^q_{--}(\bar x,\xi,{\bf \Delta}_\perp) \nonumber\\
\label{eq:31}
& = &H^q(\bar x,\xi,{\bf \Delta}_\perp) 
-\frac{\xi^2}{1-\xi^2}\, E^q(\bar x,\xi,{\bf \Delta}_\perp),\\
\label{eq:32}
F^q_{-+}(\bar x,\xi,{\bf \Delta}_\perp) & = & 
-\left(F^q_{+-}(\bar x,\xi,{\bf \Delta}_\perp) \right)^*
= \eta\frac{\sqrt{t_0-t}}{2M} \, \frac{1}{\sqrt{1-\xi^2}}\,
E^q(\bar x,\xi,{\bf \Delta}_\perp),
\eea
where
\be
\eta=\frac{\Delta^1 + i\Delta^2}{\vert\bf{\Delta}_\perp\vert},
\label{eq:33}
\ee
and
\be
\label{eq:34}
- t_0 = \frac{4{\xi}^2M_N^2}{1-\xi^2}
\ee
is the minimal value for $-t$ at given $\xi$.

\section{The convolution model for the unpolarized GPDs}
\label{gpds_mcm}

Before deriving the convolution formulas for the GPDs in the meson 
cloud model, we need to specify our conventions for the kinematical variables.
The Fock expansion of the nucleon wavefunction in Eq.~(\ref{eq:18}) 
as well as the LCWF for 
the hadron states in Eqs.~(\ref{eq:21}), (\ref{eq:22}) and (\ref{eq:27})  
do not depend on the momentum of the hadron, but only
 on the momentum coordinates of the constituents relative to the hadron 
momentum.
This fact reflects the well known result in light-front dynamics that the 
centre of mass motion can be separated from the relative motion of the 
constituents.
On the other hand, the arguments of the LCWF can most easily identified in 
reference frames 
where the hadron has zero transverse momentum.  
We call such frames ``hadron frames'', and more specifically
we introduce the names ``hadron-in'' and ``hadron-out'' for the frames 
where 
the incoming and outgoing hadron has zero transverse momentum, respectively~\cite{DFJK}.  
In general, we  denote the momenta of constituents belonging to the 
incoming hadrons with unprimed, and the momenta of constituents 
belonging to the outgoing hadron with primed variables.
Furthermore, we label quantities in the hadron-in (hadron-out) 
frame with an additional tilde (hat). 
The relations between the momenta of the constituents 
in the ``average frame'' defined 
in Eq.~(\ref{eq:28}) and the variables in the hadron frames 
are obtained via a transverse boost, i.e. a transformation that leaves the 
plus component of any four-vector $z$ unchanged.
It reads
\begin{eqnarray}
[z^+,z^-,{\bf z}_\perp]\rightarrow \left[z^+,z^--\frac{{\bf z}_\perp\cdot 
{\bf b}_\perp}{b^+}+\frac{z^+{\bf b}_\perp^2}{2(b^+)^2},{\bf z}_\perp-\frac{z^+}{b^+}{\bf b}_\perp\right],
\label{eq:35}
\end{eqnarray}
where the two parameters $b^+$ and $ {\bf b}_\perp$ are given by
$b^+=(1+\xi)\,\bar p_N^{\,+}$ and ${\bf b}_{\perp}=-{\bf\Delta}_\perp/2$ for
the transformation from the average frame to the hadron-in frame.
Likewise, a transverse boost with parameters $b^+=(1-\xi)\,\bar p_N^{\,+}$ and 
${\bf b}_\perp=+{\bf\Delta}_\perp/2$ leads from the average frame to the 
hadron-out frame.

In the following we will derive the convolution formulas
for the GPDs in the three different regions
 corresponding to $\xi\leq \bar x \leq 1,$ $-\xi\leq \bar x \leq \xi,$ 
and $-1\leq \bar x \leq -\xi.$


\subsection{The region $\xi \leq \bar x\leq 1$}
\label{sub1}

In this region the GPDs describe the emission of a quark from the nucleon with
 momentum fraction $\bar x +\xi$ and its reabsorption with momentum 
fraction $\bar x-\xi$. In the meson-cloud model, the virtual photon can 
hit 
either the bare nucleon $N$ or one of the higher Fock states. As a consequence,
 the DVCS amplitude can be written as the sum of two contributions
\begin{eqnarray}
F^q_{\lambda'_N\lambda_N}(\bar x,\xi,{\bf \Delta}_\perp) =
Z\,
F^{q, bare}_{\lambda'_N\lambda_N}(\bar x,\xi,{\bf \Delta}_\perp) 
+\delta F^q_{\lambda'_N\lambda_N}(\bar x,\xi,{\bf \Delta}_\perp),
\end{eqnarray}
where $F^{q, bare}$ is the contribution from the bare proton, described in 
terms of Fock states with three valence quarks, and $\delta F^q$ is the 
contribution from the $BM$ Fock components of the nucleon state, corresponding 
to five-parton configurations. This last term can further be split into 
two contributions, with the active quark belonging either to the 
baryon ($\delta F^{q/BM}$) or to the active quark inside the
 meson ($\delta F^{q/MB}$), 
i.e.
\begin{eqnarray}
\delta F^q_{\lambda'_N\lambda_N}(\bar x,\xi,{\bf \Delta}_\perp)=
\sum_{B,M}
\left[
\delta F^{q/BM}_{\lambda'_N\lambda_N}(\bar x,\xi,{\bf \Delta}_\perp)
+\delta F^{q/MB}_{\lambda'_N\lambda_N}(\bar x,\xi,{\bf \Delta_\perp})
\right].
\label{eq:fqb}
\end{eqnarray}

\begin{figure}[ht]
\begin{center}
\epsfig{file=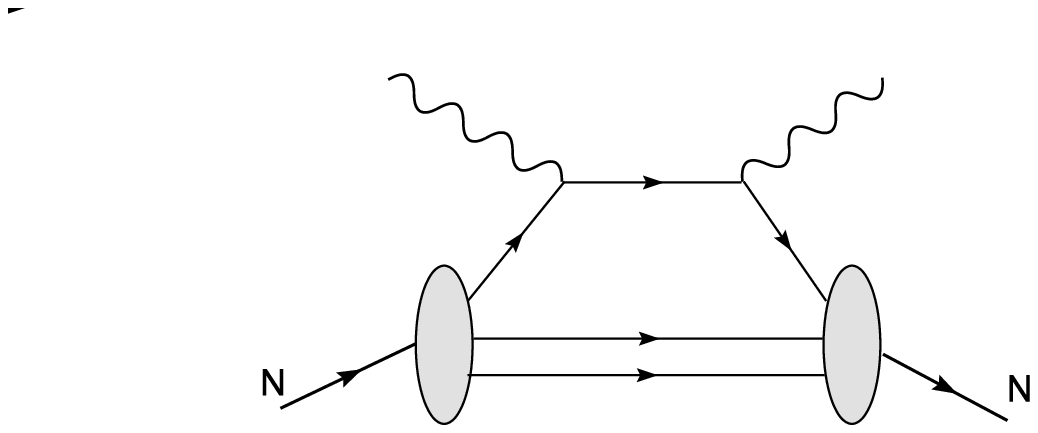,  width=7cm}
\end{center}
\caption{\small Deeply virtual Compton scattering from the bare nucleon.}
\label{fig:fig1}
\end{figure}

The valence-quark contribution corresponding to the diagram of 
Fig.~\ref{fig:fig1} can be calculated in terms of the light-front 
overlap representation derived in Ref.~\cite{DFJK}, and applied here to 
the case of $N=3$ valence quarks. It reads 
\bea
\label{eq:valence}
F^{q,bare}_{\lambda'_N\lambda_N}(\bar x,\xi,{\bf \Delta}_\perp)
&=&
\frac{1}{(1-\xi^2)}\sum_{\lambda_i\tau_i}
\sum_{j=1}^3 \delta_{s_j q}
\int[{\rm d}\bar x]_3[{\rm d}^2\bar{{\bf k}}_\perp]_3\,
\delta(\bar x-\bar x_j)
\nonumber\\
& & {}\times 
\Psi^{N,[f]\,*}_{\lambda'_N}(\{\hat x'_i, \hat{{\bf k}}'_i;
\lambda_i,\tau_i\}) 
\Psi^{N,[f]}_{\lambda_N}(\{\tilde x_i,\tilde{\bf k}_i;
\lambda_i,\tau_i\}),
\label{eq:dglapbare}
\eea
where the LCWF 
$\Psi^{N,[f]}_\lambda(\{x_i,{\bf k}_{i\perp};\lambda_i,\tau_i\})$ is the 
bare-nucleon LCWF of Eq.~(\ref{eq:18}), and $s_j$ labels the quantum numbers 
of the $j$th active parton. The integration in Eq.~(\ref{eq:dglapbare}) is over the average quark transverse momenta $\bar{\bf k}_{\perp i}$ and the average quark longitudinal momentum fractions $\bar x_i = \bar k^+_i/\bar p^+_N$.
The kinematical variables appearing as arguments in the LCWFs
  in Eq.~(\ref{eq:dglapbare}) have been defined in the ``hadron-in'' 
and ``hadron-out''  frames following the conventions of Ref.~\cite{DFJK}.
In particular, the momenta of the partons belonging to the incoming hadron
are given by
\be
\begin{array}{ll}
\displaystyle\tilde x_i= \frac{\bar x_i}{1+\xi} \eqcm \qquad& \displaystyle
\tilde{\bf k}_{\perp i}=\bar{\bf k}_{\perp i}
                        +\frac{\bar x_i}{1+\xi}\,
                         \frac{{\bf\Delta}_\perp}{2} \eqcm
\end{array}
\label{eq:tilde-args1}
\ee
for the spectator quarks ($i\neq j$);
\be
\begin{array}{ll}
\displaystyle\tilde x_j=\frac{\bar x_j+\xi}{1+\xi} \eqcm \qquad& \displaystyle
\tilde{\bf k}_{\perp j}=\bar{\bf k}_{\perp j}
                        -\frac{1-\bar x_j}{1+\xi}\,
                         \frac{{\bf\Delta}_\perp}{2}
\eqcm
\end{array} 
\label{eq:tilde-args2}
\ee
for the active quark.
Likewise, the LCWF arguments for the outgoing hadron are explicitly given by
\be
\begin{array}{ll}
\displaystyle\hat x'_i= \frac{\bar x_i}{1-\xi} \eqcm \qquad& \displaystyle
\hat{\bf k}'_{\perp i}=\bar{\bf k}^\adj_{\perp i}
                        -\frac{\bar x_i}{1-\xi}\,
                         \frac{{\bf\Delta}_\perp}{2} \eqcm
\end{array} 
\label{eq:hat-args1}
\ee
for the spectator quarks ($i\neq j$);
\be
\begin{array}{ll}
\displaystyle\hat x'_j=\frac{\bar x_j-\xi}{1-\xi} \eqcm \qquad& \displaystyle
\hat{\bf k}'_{\perp j}=\bar{\bf k}^\adj_{\perp j}
                        +\frac{1-\bar x_j}{1-\xi}\,
                         \frac{{\bf\Delta}_\perp}{2}
\eqcm
\end{array} 
\label{eq:hat-args2}
\ee
for the active quark.

\begin{figure}[ht]
\begin{center}
\epsfig{file=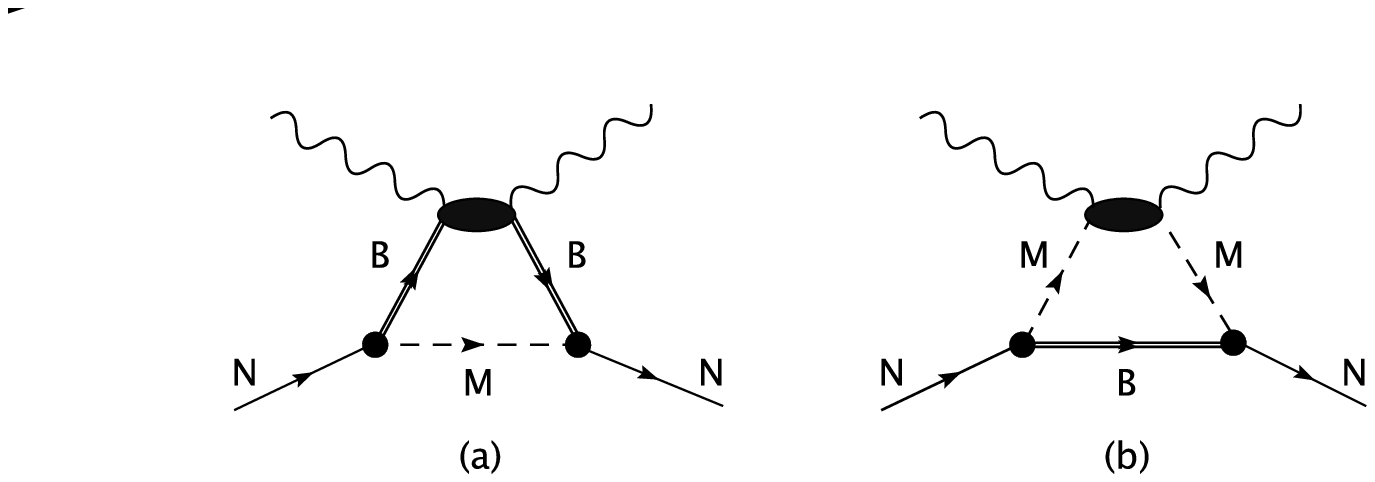,  width=12cm}
\end{center}
\caption{\small Deeply virtual Compton scattering from the virtual (a)
 baryon and (b) meson components of a dressed nucleon.}
\label{fig:fig2}
\end{figure}

We now consider the contribution from the $BM$ component of the proton 
wavefunction, starting from the $\delta F^{q/BM}_{\lambda'_N\lambda_N}$ 
term in 
Eq.~(\ref{eq:fqb}) represented in Fig. 2a.
In this case, the baryon is taken out from the initial proton with a 
fraction $\bar y_B +\xi$ of the average plus-momentum $\bar p_N^+$,
 and after the interaction with the initial and final photons
is reinserted back into the final proton with  a fraction $\bar y_B-\xi$
 of the average plus-momentum $\bar{p}_N^{\,+}$. The transverse momentum of the
 baryon is  $\bar{{\bf p}}_{B\perp}-{\bf \Delta}_\perp/2$ before, and 
$\overline{{\bf p}}_{B\perp}+{\bf \Delta}_\perp/2$ after the scattering
 process.
The meson substate is a spectator during the whole scattering process, with
 momentum in the initial and final state equal to 
$(\overline p^+_M,-\overline{{\bf p}}_{B\perp}).$ 

The baryon contribution to the $\delta F^q_{\lambda'_N\lambda_N}$ scattering 
amplitude can be easily evaluated by calculating the matrix element in
 Eq.~(\ref{eq:29}) between the $BM$ components of
the initial and final nucleon expressed in the hadron-in and 
hadron-out frame, respectively. For the initial state, 
they are given by
\begin{eqnarray}
& &
\sum_{B,M}
\int \frac{{\rm d}\tilde y_B}{\sqrt{\tilde y_B}}
\frac{{\rm d}\tilde y_M}{\sqrt{\tilde y_M}}
\delta(1-\tilde y_B-\tilde y_M)
\int \frac{{\rm d}^2\tilde{{\bf k}}_{B\perp}
 {\rm d}^2\tilde{{\bf k}}_{M\perp}}{2(2\pi)^3}
\delta^2(\tilde{{\bf k}}_{B\perp}+\tilde{{\bf k}}_{M\perp})
\nonumber\\
& &{}\times \sum_{\lambda',\lambda''}\phi^{\lambda_N\,(N,BM)}_{\lambda'\lambda''}
(\tilde y_B,\tilde{{\bf k}}_{B\perp})\,
|(\overline y_B+\xi)\,\overline p^+_N, \overline{\bf p}_{B\perp}-
\frac{{\bf \Delta}_\perp}{2},\lambda'\rangle
|\overline p^+_M, -\overline{\bf p}_{B\perp},\lambda'' \rangle.
\label{eq:43}
\end{eqnarray}

The coordinates of the baryon and meson states in the average frame are 
related to the variables in the hadron-in frame
through the transverse boost in Eq.~(\ref{eq:35}) as explained before.
They are explicitly given by
 \begin{eqnarray}
\tilde y_B= \frac{\overline y_B+\xi}{1+\xi}\eqcm &\qquad& 
\tilde{{\bf k}}_{B\perp}=\overline{{\bf p}}_{B\perp}
-\frac{1-\overline y_B}{1+\xi}
\frac{{\bf \Delta}_\perp}{2},
\label{eq:44}
\end{eqnarray}
\begin{eqnarray}
\tilde y_M=\frac{\overline y_M}{1+\xi} \eqcm &\qquad& 
\tilde{{\bf k}}_{M\perp}=\overline{{\bf p}}_{M\perp}+\frac{\overline y_M}{1+\xi}
\frac{{\bf \Delta}_\perp}{2}.
\end{eqnarray}
\noindent
Analogously, the $BM$ components of the final nucleon state are given by
\begin{eqnarray}
& &\sum_{B,M}
\int \frac{{\rm d}\hat y'_B}{\sqrt{\hat y'_B}}
\frac{{\rm d}\hat y'_M}{\sqrt{\hat y'_M}}
\delta(1-\hat y'_B-\hat y'_M)
\int \frac{{\rm d}^2\hat{\bf k}'_{B\perp}
 {\rm d}^2\hat{\bf k}'_{M\perp}}{2(2\pi)^3}
\delta^2(\hat{\bf k}'_{B\perp}+\hat{\bf k}'_{M\perp})
\nonumber\\
& &{}\times
\sum_{\lambda',\lambda''}\phi^{\lambda'_N\,(N,BM)}_{\lambda'\lambda''}
(\hat y'_B,\hat{\bf k}'_{B\perp})\,
|(\overline y_B-\xi)\,\overline p^+_N, \overline{\bf p}_{B\perp}+
\frac{{\bf \Delta}_\perp}{2},\lambda'\rangle
|\overline p^+_M, -\overline{\bf p}_{B\perp},\lambda'' \rangle,
\label{eq:46}
\end{eqnarray}
where the coordinates in the hadron-out frame and in the average frame are 
related by
\begin{eqnarray}
\hat y'_B=\frac{\overline y_B-\xi}{1-\xi}  \eqcm &\qquad& 
\hat{{\bf  k}}'_{B\perp}=\overline{{\bf p}}_{B\perp}
+\frac{1-\overline y_B}{1-\xi}\frac{{\bf\Delta}_\perp}{2},
\end{eqnarray}
\begin{eqnarray}
\hat y'_M=\frac{\overline y_M}{1-\xi} \eqcm &\qquad& 
\hat{{\bf k}}'_{M\perp}=\overline{{\bf p}}_{M\perp}-\frac{\overline y_M}{1-\xi}
\frac{{\bf \Delta}_\perp}{2}.
\end{eqnarray}

The final result for the baryon contribution  to the 
$\delta F^q_{\lambda'_N\lambda_N}$ scattering amplitude is
\begin{eqnarray}
\delta F^{q/BM}_{\lambda'_N\lambda_N}(\overline x,\xi,{\bf \Delta}_\perp)
&=&
\frac{1}{\sqrt{1-\xi^2}}
\sum_{M}\sum_{\lambda,\lambda',\lambda''}
\int_{\overline x}^1
\frac{{\rm d}\overline y_B}{\overline y_B}
\int
\frac{{\rm d}^2\overline{\bf p}_{B\perp}}{2(2\pi)^3}
F^{q/B}_{\lambda'\lambda}
\left(\frac{\overline x}{\overline y_B},\frac{\xi}{\overline y_B}, 
{\bf \Delta}_\perp\right)
\nonumber\\
& &{}\times
\phi^{\lambda_N\,(N,BM)}_{\lambda\lambda''}(\tilde y_B,\tilde{\bf k}_{B\perp})\,
[\phi^{\lambda'_N\,(N,BM)}_{\lambda'\lambda''}(\hat y'_B,\hat{\bf k}_{B\perp})]^*,
\label{eq:baryondglap}
\end{eqnarray}
\hspace{0.5cm}
where
\begin{eqnarray}
F^{q/B}_{\lambda'\lambda}
\left(\frac{\overline x}{\overline y_B},\frac{\xi}{\overline y_B},
{\bf \Delta}_\perp\right)
&=&
\frac{1}{2\sqrt{1-\xi^2/\overline y_B^2}}\int \frac{dz^-}{2\pi}\, e^{i
\overline p^+_Bz^-\,\overline x/\overline y_B\,}
\nonumber\\
& &{}\times\left.
\bra{p'^+_B,{\bf p}'_{B\perp},\lambda'}
\overline\psi(-\frac{z}{2})\,\gamma^+\, 
\psi(\frac{z}{2})\ket{p^+_B, {\bf p}_{B\perp},\lambda}\right\vert_{z^+={\bf z}_\perp=0}
\label{eq:baryionsa}
\end{eqnarray}
is the scattering amplitude from the active baryon in the $BM$ component of 
the nucleon.
The overlap representation of  $F^{q/B}_{\lambda\lambda'}$ in terms
of LCWFs is given explicitly in App.~\ref{appendixb}.

Analogously, we can derive the meson contribution to the scattering amplitude,
 corresponding to the case when the pion takes part to the interaction 
process while the baryon remains as a spectator (see Fig. 2b). 
In such a case the role of the meson and baryon substates is interchanged with 
respect to the situation described before.
The meson is taken out from the initial proton with a 
fraction $\overline y_M +\xi$ of the average plus-momentum $\overline p_N^+$,
 and after the interaction with the initial and final photons
is reinserted back into the final proton with  a fraction $\overline y_M-\xi$
 of the average plus-momentum $\overline{p}_N^{\,+}$. The transverse momentum of the
 meson is  $\overline{{\bf p}}_{M\perp}-{\bf \Delta}_\perp/2$ before, and 
$\overline{{\bf p}}_{M\perp}+{\bf \Delta}_\perp/2$ after the scattering
 process.
Viceversa, the baryon substate is inert during the whole scattering process, 
with the same momentum
$(\overline p^+_B,-\overline{{\bf p}}_{M\perp})$
in the initial and final state.
Therefore the meson contribution to the 
$\delta F^q_{\lambda'_N\lambda_N}$ scattering amplitude is
 given by
\begin{eqnarray}
\delta F^{q/MB}_{\lambda'_N\lambda_N}(\overline x,\xi,{\bf \Delta}_\perp)
&=&
\frac{1}{\sqrt{1-\xi^2}}
\sum_{B}\,
\sum_{\lambda,\lambda',\lambda''}
\int_{\overline x}^1\frac{{\rm d}\overline y_M}{\overline y_M}
\int\frac{{\rm d}^2\overline{\bf p}_{M\perp}}{2(2\pi)^3}
F^{q/M}_{\lambda'\lambda}
\left(\frac{\overline x}{\overline y_M},\frac{\xi}{\overline y_M}, {\bf \Delta}_\perp\right)
\nonumber\\
& &{}\times
\phi^{\lambda_N\,(N,BM)}_{\lambda''\lambda}(1-\tilde y_M,-\tilde{\bf k}_{M\perp})\,
[\phi^{\lambda'_N\,(N,BM)}_{\lambda''\lambda'}(1-\hat y'_M,-\hat{\bf k}_{M\perp})]^*,
\label{eq:mesondglap}
\end{eqnarray}
where  
\begin{eqnarray}
F^{q/M}_{\lambda'\lambda}
\left(\frac{\overline x}{\overline y_M},\frac{\xi}{\overline y_M},
{\bf \Delta}_\perp\right)
&=&
\frac{1}{2\sqrt{1-{\xi^2}/{\overline y_M^2}}}\int \frac{dz^-}{2\pi}\, e^{i
\overline p^+_M\,z^-\,\overline x/\overline y_M}
\nonumber\\
& &{} \times\left.
\bra{p'^+_M,{\bf p}'_{M\perp},\lambda'}
\overline\psi(-\frac{z}{2})\,\gamma^+\, 
\psi(\frac{z}{2})\ket{p_M^+, {\bf p}_{M\perp},\lambda}\right\vert_{z^+={\bf z}_\perp=0}
\label{eq:mesongpd}
\end{eqnarray}
is the scattering amplitude from the active meson in the $BM$ component of the nucleon. In Eq.~(\ref{eq:mesondglap}), the meson coordinates in the hadron-in
and hadron-out frames are related to the variables in the average frame 
by
\begin{eqnarray}
\tilde y_M=\frac{\overline y_M+\xi}{1+\xi}\eqcm &\qquad& 
\tilde{{\bf k}}_{M\perp}=\overline{{\bf p}}_{M\perp}
-\frac{1-\overline y_M}{1+\xi}
\frac{{\bf \Delta}_\perp}{2},\label{eq:53}\\
\hat y'_M=\frac{\overline y_M-\xi}{1-\xi}\eqcm &\qquad& 
\hat{{\bf  k}}'_{M\perp}=\overline{{\bf p}}_{M\perp}
+\frac{1-\overline y_M}{1-\xi}\frac{{\bf\Delta}_\perp}{2}.
\label{eq:54}
\end{eqnarray}

The explicit expression of $F^{q/M}_{\lambda\lambda'}$ in terms of LCWFs is 
given in App.~\ref{appendixc}.

\hspace{0.5cm}

\noindent {\it {\bf Forward limit}}

\hspace{0.5cm}

In the limit $\Delta^\mu\rightarrow 0$, where $\overline x$ goes over to the parton momentum fraction $x$, and $\xi={\bf \Delta}_\perp=0$, the scattering amplitude without nucleon helicity flip reduces to the ordinary parton distribution, i.e.
\begin{eqnarray}
F^q_{++}(\overline x,0,0) =
F^q_{--}(\overline x,0,0)=q(x)=
Z \,q^{bare}(x)+\delta q(x).
\end{eqnarray}

In Ref.~\cite{BPT}, we derived the forward limit of the valence-quark contribution, i.e. $F^{q, bare}_{++}(\overline x,0,0)=q^{bare}(x).$ Here we show that 
the forward limit of Eqs.~(\ref{eq:baryondglap}) and 
(\ref{eq:mesondglap}) gives the $\delta q(x)$ contribution to the parton distribution considered 
in Ref.~\cite{Speth98} within the meson-cloud model.

In the case of the active baryon, we have $\tilde y_B=\hat y'_B=y_B$ and $\tilde {\bf k}_{B\perp}=
\hat{\bf k}'_{B\perp}={\bf k}_{B\perp}$, and Eq.~(\ref{eq:baryondglap}) reduces to the following expression
\begin{eqnarray}
\delta F^{q/BM}_{++}(x,0,0)&=&
\sum_M
\int_x^1
{\rm d}y_B
\int
\frac{
{\rm d}^2
{\bf k}_{B\perp}}{2(2\pi)^3}
\frac{1}{y_B}
\sum_{\lambda,\lambda''}
|\phi_{\lambda\lambda''}^{+\,(N,BM)}(y_B,{\bf k}_{B\perp})|^2\,
F^{q/B}_{++}\left
(\frac{x}{y_B},0,0\right)\nonumber\\
&=&
\sum_M
\int_x^1\frac{{\rm d}y_B}{y_B}
q_B\left(\frac{x}{y_B}\right)\, f_{BM,N}(y_B),
\end{eqnarray}
where the splitting function
\begin{eqnarray}
f_{BM,N}(y_B)&=&\int
\frac{{\rm d}^2{\bf k}_{B\perp}}{2(2\pi)^3}
\sum_{\lambda,\lambda''} |\phi^{+\,(N,BM)}_{\lambda\lambda''}
(y_B,{\bf k}_{B\perp})|^2
\label{eq:57}
\end{eqnarray}
coincides with the definition given in Ref.~\cite{Speth98}.

Analogously, in the case of the active meson we have 
$\tilde y_M=\hat y'_M=y_M$ and 
$\tilde {\bf k}_{M\perp}= \hat{\bf k}'_{M\perp}={\bf k}_{M\perp}$ and 
Eq.~(\ref{eq:mesondglap}) reduces to the following expression
\begin{eqnarray}
\delta F^{q/MB}_{++}(x,0,0)&=&
\sum_B\,
\sum_{\lambda,\lambda''}
\int_x^1
\frac{{\rm d}y_M}{y_M}
\int
\frac{{\rm d}^2{\bf k}_{M\perp}}{2(2\pi)^3}
\nonumber\\
& &{}\times
|\phi_{\lambda\lambda''}^{+\,(N,BM)}(1-y_M,-{\bf k}_{M\perp})|^2\,
F^{q/MB}_{++}
\left(\frac{x}{y_M},0,0\right)
\nonumber\\
&=&
\sum_B
\int_x^1\frac{{\rm d}y_M}{y_M}f_{MB,N}(y_M)
q_M\left(\frac{x}{y_M}\right)\, ,
\label{eq:58}
\end{eqnarray}
where we use the definition
\begin{eqnarray}
f_{MB,N}(y_B)&=&\int
\frac{{\rm d}^2{\bf k}_{B\perp}}{2(2\pi)^3}
\sum_{\lambda,\lambda''} |\phi^{+\,(N,BM)}_{\lambda\lambda''}
(1-y_M,-{\bf k}_{M\perp})|^2.
\label{eq:59}
\end{eqnarray}

Since the probability for the dressed nucleon to consist of a bare baryon 
and meson is independent of which one interacts with the probe,
we assumed in Eq.~(\ref{eq:59}) the following condition
\be
f_{MB,N}(y_M)= f_{BM,N}(1-y_M).
\label{eq:60}
\ee
This property cannot be derived a priori from the definition of the
 convolution model, but it is an additional physical input which may be used to
 restrict the vertex functions of the model as we will discuss in 
Sect.\ref{model}.
Moreover, the relation~(\ref{eq:60}) automatically ensures
both momentum and baryon number sum rules.

As a final result, the contribution of the higher Fock states to the parton distribution is given by
\begin{equation}
 \delta q(x) = \sum_{MB} \left [
  \int_x^1 \frac{dy}{y}\,f_{MB/N}(y)\, q_M \left(\frac{x}{y}\right)
 +  \int_x^1 \frac{dy}{y}\, f_{BM/N}(y)\, q_B \left(\frac{x}{y}\right)
 \right ] ,
\label{eq:splitting}
\end{equation}
which coincides with the formulation of the meson cloud model in deep 
inelastic process (see, for example, Ref.~\cite{Speth98} and reference 
therein). 


\subsection{The region $-1 \leq \overline x\leq -\xi$}
\label{sub2}

In this region, the scattering amplitude describes the emission of an 
antiquark from the nucleon with momentum fraction $-(\overline x+\xi)$ and 
its reabsorption with momentum fraction $-(\overline x-\xi)$. As a consequence, 
the only non vanishing contribution can come from the active antiquark in the
meson substate of the $BM$ Fock component of the nucleon wavefunction, i.e.
\begin{eqnarray}
F^q_{\lambda'_N\lambda_N}(\overline x,\xi,{\bf \Delta}_\perp)
=\delta F^{q/MB}_{\lambda'_N\lambda_N}(\overline x,\xi,{\bf \Delta}_\perp),
\end{eqnarray}
where $\delta F^{q/MB}_{\lambda'_N\lambda_N}(\overline x,\xi,{\bf \Delta}_\perp)$
corresponds to the meson scattering amplitude illustrated in 
Fig. 2(b), and is given by
\begin{eqnarray}
\delta F^{q/MB}_{\lambda'_N\lambda_N}(\overline x,\xi,{\bf \Delta}_\perp)
&=&
\frac{1}{\sqrt{1-\xi^2}}
\sum_{B}\,
\sum_{\lambda,\lambda',\lambda''}
\int_{-\overline x}^1
\frac{{\rm d}\overline y_M}{\overline y_M}
\int
\frac{{\rm d}^2\overline{\bf p}_{M\perp}}{2(2\pi)^3}
F^{q/M}_{\lambda'\lambda}
\left(\frac{\overline x}{\overline y_M},\frac{\xi}{\overline y_M}, {\bf \Delta}_\perp\right)
\nonumber\\
& &{}\times
\phi^{\lambda_N\,(N,BM)}_{\lambda''\lambda}(1-\tilde y_M,-\tilde{\bf k}_{M\perp})\,
[\phi^{\lambda'_N\,(N,BM)}_{\lambda''\lambda'}(1-\hat y'_M,-\hat{\bf k}_{M\perp})]^*.
\label{eq:mesondglap2}
\end{eqnarray}
Here the relations between the coordinates in the 
hadron frames and in the average frame are the same as in Eqs.~(\ref{eq:53})
and (\ref{eq:54}).
In addition,
we note that Eq.~(\ref{eq:mesondglap2}) corresponds to the same convolution 
formula~(\ref{eq:mesondglap}), with the integration range over $\overline y_M$ 
between $-\overline x$ and $1$, and with the explicit LCWF overlap 
representation of $F^{q/M}_{\lambda'\lambda}$ in the range 
$-1 \leq \overline x\leq -\xi$ given in App.~\ref{appendixc}.

\hspace{0.5cm}


\noindent {\it {\bf Forward limit}}

\hspace{0.5cm}

The scattering amplitude from the meson substate has the following forward limit:
\begin{eqnarray}
F^{q/M}_{++}(\frac{x}{y_M},0,0)=
q(\frac{x}{y_M})=-\bar q(-\frac{x}{y_M}).
\end{eqnarray}
As a consequence, in the forward limit Eq.~(\ref{eq:mesondglap2}) reduces to
\begin{eqnarray}
\delta F^{q/MB}_{++}(x,0,0)&=&
-
\sum_B
\int_{-x}^1\frac{{\rm d}y_M}{y_M}
\bar q_M\left(-\frac{x}{y_M}\right)\, f_{MB,N}(y_M)\nonumber\\
&=&-\delta\bar q(-x).
\end{eqnarray}
 

\subsection{The region $-\xi \leq \overline x\leq \xi$}
\label{sub3}

\begin{figure}[ht]
\begin{center}
\epsfig{file=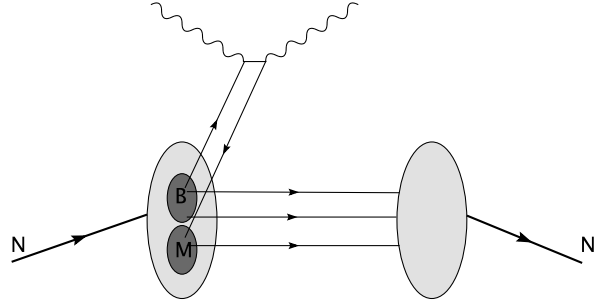,  width=8cm}
\end{center}
\caption{\small Deeply virtual Compton scattering in the ERBL region.}
\label{fig:fig3}
\end{figure}

In this region, the scattering amplitude describes the emission of a 
quark-antiquark pair from the initial proton. As discussed in Ref.~\cite{DFJK},
 in the Fock-state decomposition of the initial and final nucleons we have to
 consider only terms where the initial state has the same parton content as the
 final state plus an additional quark-antiquark pair. In the present 
meson-cloud model, the initial state is given by the five-parton component in
 Eq.~(\ref{eq:27}), while the final state is described by the three-valence
 quark configuration given in Eq.~(\ref{eq:18}), multiplied by the 
normalization factor $\sqrt{Z}$ . To label the coordinates of the initial and 
final partons we follow the conventions explained in Ref.~\cite{DFJK}. This 
means that we use the same numbering for the spectator partons in the LCWFs of
 the initial and final state partons. Thus the three partons in the outgoing
 nucleon are numbered not as $i=1,2,3,$
but as $i=1,...,5$ with the labels of the active quark $(j)$ and the active 
antiquark $(j')$ omitted. 
In principle, we can distinguish between two cases: $i$) 
the active quark belongs to the baryon substate ($j=1,2,3$), and the active 
antiquark is in the parton configuration of the meson substate of the initial
 nucleon; $ii$) both the active quark and  antiquark belong to the meson substate
 of the initial nucleon ($j=4$ and $j'=5$) and the baryon is a spectator
during the scattering process.
However, this last contribution is vanishing because it involves 
the overlap of two orthogonal states, i.e. the wave 
functions of the baryon in the initial state and of the bare 
nucleon in the final state.
As a consequence, 
the only non vanishing contribution corresponds to the case $i$) 
which is pictured in Fig.~\ref{fig:fig3}.

The final result for the scattering amplitude in the ERBL region is 
derived using the overlap representation of Ref.~\cite{DFJK} with
the initial wavefunction describing the initial 5-parton configuration 
given in Eq.~(\ref{eq:27}) and the final state 
described by the valence-quark wavefunction in Eq.~(\ref{eq:18}). 
It  reads

\begin{eqnarray}
F^{q}_{\lambda'_N\lambda_N}&=&
\frac{\sqrt{Z}}{(1-\xi)\,(1+\xi)^2}
\sum_{B,M}\,\sum_{\lambda_i,\tau_i}
\sum_{j=1}^3\sum_{j'}\delta_{s_j,s_q}
\delta_{\lambda_j,\,-\lambda_{j'}}\,
K_c\,
\int{\rm d}\tilde y_B\int{\rm d}\tilde y_M
\delta(1-\tilde y_B-\tilde y_M)\nonumber\\
& &{}\times
\int{\rm d}^2\tilde {\bf k}_{B\perp}
\int{\rm d}^2\tilde {\bf k}_{M\perp} 
\delta(\tilde {\bf k}_{B\perp}+\tilde {\bf k}_{M\perp})
\, \int {\rm d}\overline x_j\int 
\prod_{i=1\atop i\ne j,j'}
{\rm d}\overline x_i
\int {\rm d}^2\overline{\bf k}_{j\perp}\int 
\prod_{i=1 \atop i\ne j,j'}^5{\rm d}^2\overline{\bf k}_{i\perp}
\nonumber\\
& &{}\times
\,\frac{1}{[2(2\pi)^3]^3} 
\,\delta(\overline x_j-\overline x)
\,\delta\left(\tilde y_B-\sum_{i=1\atop i \ne j}^3\frac{\overline x_i}{1+\xi}
-\frac{\overline x_j+\xi}{1+\xi}\right)\,
\delta \left(1-\xi-\sum_{i=1\atop i\ne j,j'}^5 \overline x_i\right)\nonumber\\
& &{}\times
\delta^2 \left
(\tilde{\bf k}_{B\perp}-\sum_{i=1}^3\overline{\bf k}_{i\perp}
+\frac{{\bf \Delta}_\perp}{2}(1-\tilde y_B)
\right)\,
\delta^2 \left
(\frac{{\bf \Delta}_\perp}{2}-\sum_{i=1\atop \ne j,j'}^5\overline{\bf k}_{i\perp}
\right)\nonumber\\
& &{}\times
\tilde \Psi^{5q,[f]}_{\lambda_N}(\tilde y_B,\tilde{\bf k}_{B\perp};
\{\tilde x_i,\tilde{\bf k}_{i\perp},\lambda_i,\tau_i\}_{i=1,...,5})
[\tilde \Psi^{3q,[f]}_{\lambda'_N}
(\{\hat x'_i,\hat{\bf k}'_{i\perp},\lambda_i,\tau_i\}_{i=1,2,3})]^*,
\label{eq:66}
\end{eqnarray}
where $K_c$ is a color factor which comes from the color component
of the initial- and final-state wavefunction.
By taking the color component equal to 
$\sum_{ijk}1/\sqrt{6}\,\varepsilon_{ijk}\,|q^iq^jq^k\,\rangle$ and 
 $\sum_{ij}1/\sqrt{3}\,\delta_{ij}|q^iq^j\,\rangle$ for the baryon and meson 
state, respectively, we have $K_c=1/\sqrt{3}$.

In Eq.~(\ref{eq:66}) the coordinates of the initial partons in the 
hadron-in frame are related to the parton momenta in the average frame by
\be
\begin{array}{ll}
\displaystyle
\tilde x_i= \frac{\overline x_i}{1+\xi} \eqcm \qquad& 
\displaystyle
\tilde{\bf k}_{\perp i}=\overline{\bf k}_{\perp i}
                        +\frac{\overline x_i}{1+\xi}\,
                         \frac{{\bf\Delta}^\adj_\perp}{2} \eqcm
\qquad \qquad \mbox{for }i\neq j,j' 
\eqcm\\
\displaystyle
\tilde x_j=\frac{\overline x_j+\xi}{1+\xi} \eqcm \qquad& 
\displaystyle
\tilde{\bf k}_{\perp j}=\overline{\bf k}_{\perp j}
                        -\frac{1-\overline x_j}{1+\xi}\,
                         \frac{{\bf\Delta}^\adj_\perp}{2}
\eqcm\\
\displaystyle
\tilde x_{j'}=-\frac{\overline x_j-\xi}{1+\xi} \eqcm \qquad& 
\displaystyle
\tilde{\bf k}_{\perp j'}=-\overline{\bf k}_{\perp j}
                        -\frac{1+\overline x_j}{1+\xi}\,
                         \frac{{\bf\Delta}^\adj_\perp}{2},
\end{array}
\label{eq:67}
\ee
where for the average momentum of the spectators partons  we used the standard
definition $  \overline k_i=\frac{1}{2}\, (k_i+k'_{i})$, while 
for the active quark we defined  $\overline k_j=\frac{1}{2}\, (k_j-k'_{j}),$
i.e. half the relative momentum between the active quark and antiquark.
The corresponding relations for the coordinates of the final partons in the 
hadron-out and average frames are
\begin{eqnarray} 
\label{eq:hat-args-central}
\hat x'_i= \frac{\overline x_i}{1-\xi} \eqcm &\qquad& 
\hat{\bf k}'_{\perp i}=\overline{\bf k}^\adj_{\perp i}
                        -\frac{\overline x_i}{1-\xi}\,
                         \frac{{\bf\Delta}^\adj_\perp}{2}
\eqcm
\qquad \qquad \mbox{for }i\neq j,j' 
\eqpt
\label{eq:68}
\end{eqnarray} 

Finally, the five-parton component of the initial 
state wavefunction $\tilde \Psi^{5q,[f]}_{\lambda_N}$ in Eq.~(\ref{eq:66})
is given by
\begin{eqnarray}
& & \tilde \Psi^{5q,[f]}_{\lambda_N}(\tilde y_B,\tilde{\bf k}_{B\perp};
\{\tilde x_i,\tilde{\bf k}_{i\perp},\lambda_i,\tau_i\}_{i=1,...,5})\nn\\
&&{}\quad=\frac{1}{\tilde y_B^{3/2}}\frac{1}{\tilde y_M}
\sum_{\lambda',\lambda''}
\frac{V^{\lambda_N}_{\lambda',\lambda''}(N,BM)}
{M^2_N-M^2_{BM}(\tilde y_B,\tilde {\bf k}_{B\perp})}
\nn\\
& & {}\qquad\times
\tilde \Psi_{\lambda'}^{B,\,[f]}(\{\tilde 
\zeta_i,\tilde {\bold \kappa}_{i\perp},\lambda_i,\tau_i\}_{i=1,2,3})
\tilde \Psi_{\lambda''}^{M,\,[f]}(\{\tilde \zeta_i,\tilde
{\bold \kappa}_{i\perp},\lambda_i,\tau_i\}_{i=4,5})
\label{eq:5-3wf}
\end{eqnarray}
with the coordinates in the 
baryon-in ($\zeta_i,$ with $i=1,2,3$) and meson-in ($\zeta_i,$ with $i=4,5$) 
frames 
defined as
\be
\begin{array}{ll}
\displaystyle
\tilde \zeta_i =\frac{\tilde x_i}{\tilde y_B} \eqcm \qquad&  
\tilde{\bold \kappa}_{i\perp}=\tilde{\bf k}_{i\perp}-\tilde\zeta_i\tilde{\bf k}_{B\perp},\quad i=1,2,3,\\
\displaystyle
\tilde \zeta_i =\frac{\tilde x_i}{\tilde y_M} \eqcm \qquad&  
\tilde{\bold \kappa}_{i\perp}=\tilde{\bf k}_{i\perp}-\tilde\zeta_i\tilde{\bf k}_{M\perp},\quad i=4,5.
\end{array}
\label{eq:70}
\ee


\section{Model for the pion cloud of the proton}
\label{model}

In this section  we specify the ingredients for the model calculation
of the unpolarized GPDs of the proton.
We restrict ourselves to consider only 
the pion-cloud contribution disregarding the contributions from mesons 
of higher masses which are suppressed.
As a consequence, the accompanying baryon in the $\ket{B\pi}$ component of
the dressed proton is a nucleon or a $\Delta.$

\hspace{0.5cm}

\noindent{\it Vertex functions}

\hspace{0.5cm}

The vertex functions for the transition 
$p\rightarrow B \pi$ are given by

\begin{eqnarray}
V^{\lambda}_{\lambda',0}(p,N\pi) &=&  i\,  g_{pN\pi}\, 
\bar u_{\lambda'}(\tilde p'_N)\gamma_5u_\lambda(\tilde p_p),\nn\\
V^{\lambda}_{\lambda',0}(p, \Delta\pi) &=& 
i \, g_{p \Delta \pi }\,\bar u_{\lambda'}^\mu(\tilde p'_\Delta)\, 
(p_p-p'_\Delta)_\mu
u_\lambda(\tilde p_p)
\label{eq:71}
\end{eqnarray}
where $ u_\lambda(\tilde p_N)$ and 
$u^\mu_{\lambda'}(\tilde p_\Delta)$ are the nucleon spinor and 
 the Rarita-Schwinger spinor, respectively, defined 
in Appendix~\ref{appendixa}.

Because of the extended nature of the vertices one has to replace the 
coupling constants in Eq.~(\ref{eq:71}) with
phenomenological vertex form factors, $G_{NBM}(y,k^2_\perp )$, which 
parametrize the unknown dynamics at the vertices.
We use the following parametrization~\cite{MST99}
\begin{equation} 
G_{NBM}(y,k_\perp^2 )= g_{NBM}
\left( \frac{\Lambda^2_{BM}+ m_N^2}{
\Lambda^2_{BM}+{M}^2_{BM}(y,k_\perp^2)}\right)^2 ,
\label{eq:74}
\end{equation}
where $\Lambda_{BM}$ is the cut-off parameter and the invariant mass of the
 baryon-meson fluctuation ${M}^2_{BM}$ is given in Eq.~(\ref{eq:11}).
We note that the dipole parametrization of the vertex form factors satisfies 
the condition
$G_{NBM}(y,k_\perp^2 )= G_{NBM}(1-y,k_\perp^2 )$ which automatically 
guarantees the property (\ref{eq:60}) for the splitting functions.

For the coupling constants we use the numerical values given 
in Refs.~\cite{MST99,BoTho99},
i.e. $g^2_{NN\pi}/4\pi=13.6$ and $g^2_{N \Delta\pi}/4\pi = 11.08$ GeV$^{-2}$, 
with $g_{NN\pi}=g_{pp\pi^0}$ and $g_{N\Delta\pi}=g_{p\Delta^{++}\pi^-}$. 
The coupling of a given type of transition with different isospin components 
is obtained in terms of isospin Clesch-Gordan coefficients, i.e.
$g_{pn\pi^+}=- \sqrt{2}g_{pp\pi^0}$, 
$g_{p\Delta^0\pi^+}
= -g_{p\Delta^+\pi^0}/\sqrt{2}
= g_{p\Delta^{++}\pi^-}/\sqrt{3}$.

The violation of the Gottfried sum rule and flavor symmetry puts constraints 
on the magnitude of the cut-off parameters. We use the values 
$\Lambda_{BM} =1.0$ GeV and $\Lambda_{BM} =1.3$ GeV for the $\pi N$ 
and $\pi\Delta$ components, respectively, because they give contributions
 to the $\bar u$ and
$\bar d$ which are consistent with the requirement that the meson-cloud 
component of the sea-quark contribution cannot be larger than the measured
 sea quark and also with flavor-symmetry violation~\cite{MST99}. 

With the specified parameters, in the case of the 
{$p\rightarrow$ $B\pi$} transition one has~\cite{BoTho99},
\begin{eqnarray*}
& & P_{N \pi /p}=P_{p \pi^0 /p}+P_{n\pi^+ /p}=3P_{p \pi^0 /p}= 13\%,\\ 
& & P_{\Delta \pi/p}=P_{\Delta^{++} \pi^-/p}+P_{\Delta^+ \pi^0/p}
+P_{\Delta^0 \pi^+/p}=2P_{\Delta^{++} \pi^-/p}=11\%.
\end{eqnarray*}

\hspace{0.5cm}

\noindent{\it LCWFs in the constituent quark model}

\hspace{0.5cm}

As explained in details in Ref.~\cite{BPT},
the LCWFs of the baryons and pion in Eqs.~(\ref{eq:21}) and (\ref{eq:22}), 
respectively, can be expressed in terms of the canonical wavefunctions,
solutions of the instant-form Hamiltonian for the $N$-valence quarks 
of the hadron, through the following relation
\bea
\label{eq:psifc}
 \Psi_\lambda^{[f]}
(\{x_i,{\bf k}_{\perp i};\lambda_i,\tau_i\}_{i=,1,..,N})  
& = &2(2\pi)^3\frac{1}{\sqrt{M_0}}
\prod_{i=1}^N
\left(\frac{\omega_i}
{x_i}\right)^{1/2}\nn\\
& &\times
\sum_{\mu_1,...,\mu_N}
\Psi_\lambda^{[c]}(\{{\bf k}_i;\mu_i,\tau_i,\mu_i\}_{i=1,...,N})
\prod_{i=1}^N {D}^{1/2\,*}_{\mu_i\lambda_i}(R_{cf}({ \tilde k}_i)),
\label{eq:75}
\eea 
where $\Psi_\lambda^{[c]}$ is the canonical wavefunction,  
and ${D}^{1/2\,*}_{\mu_i\lambda_i}(R_{cf}({\tilde k}_i))$ are the Melosh
 rotations defined in Ref.~\cite{BPT}.
In Eq.~(\ref{eq:75}), $\omega_i=\sqrt{m_i^2+{\bf k}_i^2}$ is the energy
of the $i$-th quark, and $M_0=\sum_i \,\omega_i$ is the free mass
of the system of $N$ non-interacting quarks.

In the case of the nucleon, we adopt the canonical wavefunction 
of the relativistic CQM model of Ref.~\cite{FTV99}, 
given by a product of a space and a spin-isospin
term which is SU(6) symmetric. 
For a more detailed discussion of the model see Ref.~\cite{FTV99}.

The $\Delta$ is described as a state of isospin $T=3/2$ obtained as 
a pure spin-flip excitation of the nucleon, 
with the corresponding spatial part of the wavefunction 
equal to that of the nucleon.

Finally, the canonical wavefunction of the pion is taken 
from ref.~\cite{Choi99} and reads 
\begin{eqnarray}
\Psi^{\pi,[c]}(\vec{k}_1,\vec{k}_2;\mu_1,\mu_2)
=\frac{i}{\pi^{3/4}\beta^{3/2}}
\left(\oneh\mu_1\oneh\mu_2|00\right) \exp{(-k^2/(2\beta^2))},
\label{eq:can_psi}
\end{eqnarray}
with $\vec k=\vec k_1=-\vec k_2$,  and
$\beta=0.3659$ GeV.
The choice of the model from Ref.~\cite{Choi99} is consistent with the
hypercentral CQM we adopt for the nucleon and the $\Delta$, 
since the central potential between
the two constituent quarks in the pion is described as a linear 
confining term plus
Coulomb-like interaction. As explained 
in more details in Ref.~\cite{Choi99}, the canonical expression 
(\ref{eq:can_psi}) 
represents a variational solution to the mass equation. The phase of the pion wavefunction (\ref{eq:can_psi}) is consistent with that of the antiquark spinors of Ref.~\cite{Brodsky:1989pv}.

\section{Results}
\label{results}

The general formalism developed in the preceding sections has been applied to calculate the unpolarized GPDs $H^q$ and $E^q$ for the proton dressed by a pion cloud. The different contributions coming from the $p\rightarrow BM$ fluctuations are derived from the basic $p\rightarrow p\pi^0$ and $p\rightarrow \Delta^{++}\pi^-$ transitions. Due to the isospin relations between the different charged channels and the SU(6) symmetry of the spin-isospin part of the proton wavefunction, in the region $\xi\leq \overline x\leq 1$ the other contributions are obtained making use of the following relations:
\be
\begin{array}{ll}
\delta F^{u/n\pi^+}=2\, \delta F^{d/p\pi^0}\eqcm \qquad&
\delta F^{d/n\pi^+}=2\, \delta F^{u/p\pi^0},\\  
\delta F^{u/\Delta^{0}\pi^+}=\frac{1}{4}\,\delta F^{u/\Delta^+\pi^0}
\eqcm \qquad&
\delta F^{d/\Delta^{0}\pi^+}=\delta F^{d/\Delta^+\pi^0}\\ 
\delta F^{u/\Delta^{++}\pi^-}=\frac{9}{4}\,\delta F^{u/\Delta^+\pi^0}, &
\delta F^{u/\pi^0 p}=\delta F^{d/\pi^0 p},\\ 
\delta F^{u/\pi^+ n}=4 \delta F^{u/\pi^0 p}, & 
\delta F^{u/\pi^+ \Delta^0}= \delta F^{u/\pi^0 \Delta^+},\\ 
\delta F^{d/\pi^- \Delta^{++}}=3 \delta F^{d/\pi^0 \Delta^+}. & 
\end{array}
\ee
Similarly, in the $-1\leq \overline x\leq-\xi$ region we have the following relations:
\be
\begin{array}{ll}
\delta F^{d/\pi^+ n}=4 \delta F^{d/\pi^0 p},& 
\delta F^{d/\pi^+ \Delta^0}= \delta F^{d/\pi^0 \Delta^+},\\
\delta F^{u/\pi^- \Delta^{++}}=3 \delta F^{u/\pi^0 \Delta^+},& 
\delta F^{d/\pi^0 p}= \delta F^{u/\pi^0 p}.
\end{array}
\ee
In addition, for $\xi\le \overline x\le 1$, we have
\be
\begin{array}{ll}
\delta F^{u/\pi^0 p} (\overline x,\xi,t) &= - \delta F^{d/\pi^0 p} (-\overline x,\xi,t) ,\\
\delta F^{d/\pi^0 p} (\overline x,\xi,t) &= - \delta F^{u/\pi^0 p} (-\overline x,\xi,t) ,\\
\delta F^{d/\pi^- \Delta^{++}} (\overline x,\xi,t) &= - \delta F^{u/\pi^- \Delta^{++}} (-\overline x,\xi,t) .
\end{array}
\ee

The multidimensional integration required for the numerical computation of the different contributions to $F^q_{\lambda'_N\lambda_N}$ was implemented in a parallel computation using the parallelized version of the VEGAS routine of Ref.~\cite{Kreckel}. In this way one makes easier an otherwise time-consuming computation. The results presented in this Section have been obtained for some combinations of $t$ and $\xi$ as an example of the effects introduced by the sea.

\begin{figure}[ht]
\begin{center}
\epsfig{file=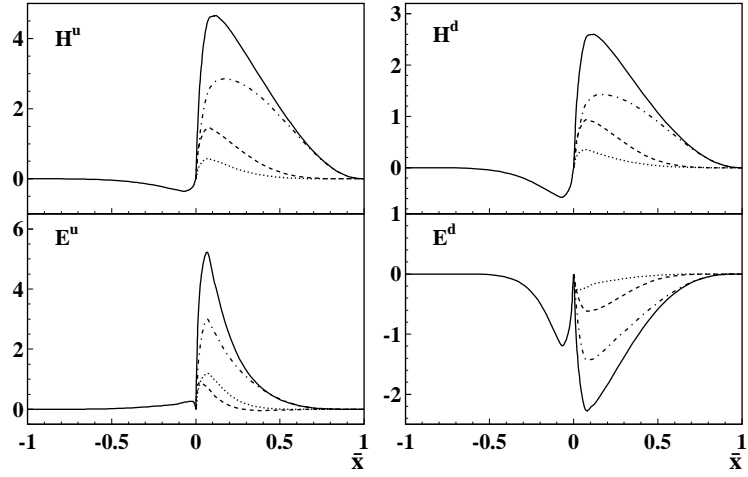,  width=10cm}
\end{center}
\caption{\small The different contributions to the spin-averaged ($H^q$, upper panels) and helicity-flip ($E^q$, lower panels) generalized parton distributions calculated in the meson-cloud model for flavours $u$ (left panels) and $d$ (right panels), at $\xi=0$ and $t=0$. Dashed lines: baryon contribution from the $\vert BM\rangle$ component. Dotted lines: meson contribution from the $\vert BM\rangle$ component.  Dashed-dotted lines: contribution from the bare nucleon. Full curves: total result as a sum of the different contributions.
}
\label{fig:fig4}
\end{figure}

First let us study the forward limit, $\xi=0$, $t=0$. 

In Fig.~\ref{fig:fig4} the spin-averaged $H^q$ and the helicity-flip $E^q$ GPDs are plotted together with the separated contributions from the bare proton (dashed-dotted line), the baryon (dashed lines) and the meson (dotted lines) in the baryon-pion fluctuation. All these contributions add up incoherently to give the total result (full curves). The bare proton contribution is the same as that already calculated in Ref.~\cite{BPT}, rescaled by the wavefunction renormalization constant $Z$. It is always positive within its support ($0\le\overline x\le 1$) with the exception of $E^d$ for which it is negative. The same behaviour characterizes the baryon contribution from the baryon-pion fluctuation that is also limited to the range $0\le\overline x\le 1$, consistently with the assumption that the only active degrees of freedom for such a baryon are the valence quarks. Both contributions vanish at the end points of their support. The sea-quark contribution, extending all over the full range $-1\le\overline x\le 1$, is determined by the antiquark residing in the meson of the baryon-pion fluctuation. The resulting effect of the pion cloud is thus to add a contribution for negative $\overline x$ and to increase the magnitude of the GPDs for positive $\overline x$ with respect to the case of the bare proton. In particular, for positive and small $\overline x$ the pion cloud contribution as a whole is comparable to that of the bare proton, confirming the important role of the sea at small $\overline x$ found within the chiral quark-soliton model~\cite{petrov,schweitzer05,wakamatsu}. Since for $\overline x>0$ the contribution of the baryon-meson fluctuation is effective only at low $\overline x$ values, the same faster fall off of $E^q$ with respect to $H^q$ for $\overline x\to 1$ is obtained as in Ref.~\cite{BPT}. This is a consequence of the decreasing role of the Melosh transform to generate angular momentum in $E^q$ with increasing quark momentum.

In all cases at $\overline x=0$ the GPDs have a zero. This is due to fact that in the overlap integrals the various terms of the proton wavefunction are taken at one of their end points. This peculiar feature was discussed in Ref.~\cite{diehlc} and explained as an artifact due to the truncated Fock-state expansion. The singular behaviour of the usual parton distributions for $\overline x\to 0$ cannot be obtained from any finite number of Fock-state contributions, all of which vanish at $\overline x=0$.

In any case, the forward limit of the first moment sum rules for the spin-averaged GPDs,
\be
\int_{-1}^1 dx \, H^u(x,0,0) = 2, \quad \int_{-1}^1 dx \, H^q(x,0,0) = 1, 
\ee
is correctly fulfilled. For the helicity-flip GPDs the first moment sum rule reads
\be
\int_{-1}^1 dx \, E^q(x,0,0) = \kappa^q,
\ee
where $\kappa^u$ ($\kappa^d$) is the anomalous magnetic moment of the $u$ ($d$) quarks, with $\kappa^u+\kappa^d=3(\kappa^p +\kappa^n)$ and $\kappa^u-\kappa^d=\kappa^p -\kappa^n$, $\kappa^p$ and $\kappa^n$ being the proton and neutron anomalous magnetic moments, respectively.

Experimentally, we have $\kappa^p=1.793$ and $\kappa^n=-1.913$, thus giving $\kappa^u=1.673$ and $\kappa^d=-2.033$.
Without the pion cloud the values of  the nucleon anomalous magnetic moments
 were found to be rather far from the experimental ones, i.e. $\kappa^p=0.91$ 
and $\kappa^n=-0.82$, corresponding to $\kappa^u=1.0$ and 
$\kappa^d=-0.74$~\cite{BPT}. This result is, however, in agreement with 
analogous light-front calculations with point-like 
quarks~\cite{cardarelli,ptb,simula} and the values derived in 
the forward limit  of GPDs derived in the Nambu-Jona-Lasinio 
model~\cite{mineo}. Including the pion cloud we have $\kappa^u=1.14$ 
and $\kappa^d=-1.03$, corresponding to $\kappa^p=1.10$ and $\kappa^n=-1.07$, 
closer to the phenomenological values. Similar effects are also expected 
for the pion-cloud contribution to the form factors at finite value of $t$, 
improving the agreement between
the CQM predictions and the experimental
results~\cite{PBprogress}.

\begin{figure}[ht]
\begin{center}
\epsfig{file=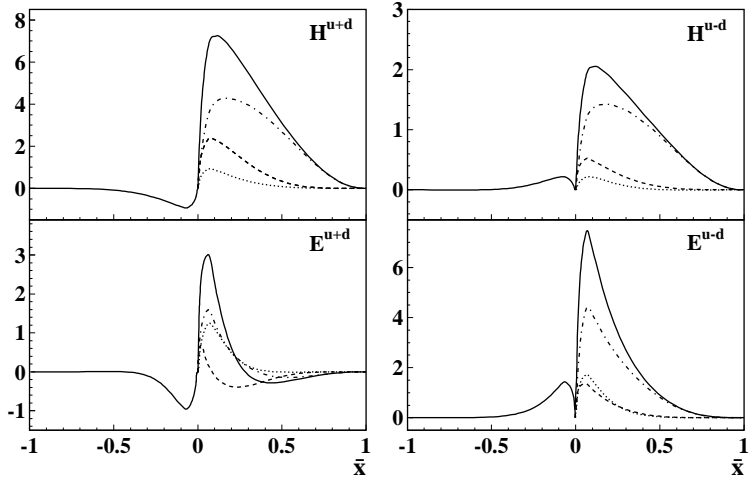,  width=10cm}
\end{center}
\caption{\small Isoscalar ($u+d$, left panels)  and isovector ($u-d$, left panels) combinations of the spin-averaged (upper panels) and helicity-flip (lower panels) generalized parton distributions calculated in the meson-cloud model, at $\xi=0$ and $t=0$. 
}
\label{fig:fig5}
\end{figure}

In Fig.~\ref{fig:fig5} the same results plotted in Fig.~\ref{fig:fig4} are reorganized to show the isoscalar $u+d$ and isovector $u-d$ combinations. In the large-$N_c$ limit $H^u+H^d$ and $E^u-E^d$ appear in leading order in $N_c$, while $H^u-H^d$ and $E^u+E^d$ appear in subleading order~\cite{goeke}. Also in our model $H^u+H^d$ and $E^u-E^d$ are found to be larger than $H^u-H^d$ and $E^u+E^d$. The behaviour of the isoscalar combinations is also very similar to that provided by the chiral quark-soliton model~\cite{petrov,schweitzer05}. The sea-quark contribution is an odd (even) function of $\overline x$ for the isoscalar (isovector) combination. As discussed above, the isovector combinations vanish at $\overline x=0$ due to the truncated Fock-state expansion of the proton wavefunction. Would the dip at $\overline x=0$ be filled by including the whole series expansion, the behaviour of the isovector combinations would resemble that derived in the chiral quark-soliton model~\cite{wakamatsu}, where GPDs in the neighbourhood of $\overline x=0$ are mostly determined by the Dirac sea. In our model the sea contribution comes only from the antiquark present in the one-pion component of the cloud.

Anyway, since there are no gluons in the model the total momentum of the proton is carried by quarks and antiquarks only. Therefore the second moment sum rule,
\be
\int_{-1}^1 dx \, x \, (H^u + H^d)(x,0,0) = M^Q = 1.
\ee
and the spin sum rule,
\be
\int_{-1}^1 dx \, x \, (H^u + H^d + E^u + E^d)(x,0,0) = 2J^Q = 1.
\ee
are consistently fulfilled in the model. 
In addition,
\be
\int_{-1}^1 dx \, x \, (E^u + E^d)(x,0,0) = 2J^Q - M^Q= 0.
\ee

\begin{figure}[ht]
\begin{center}
\epsfig{file=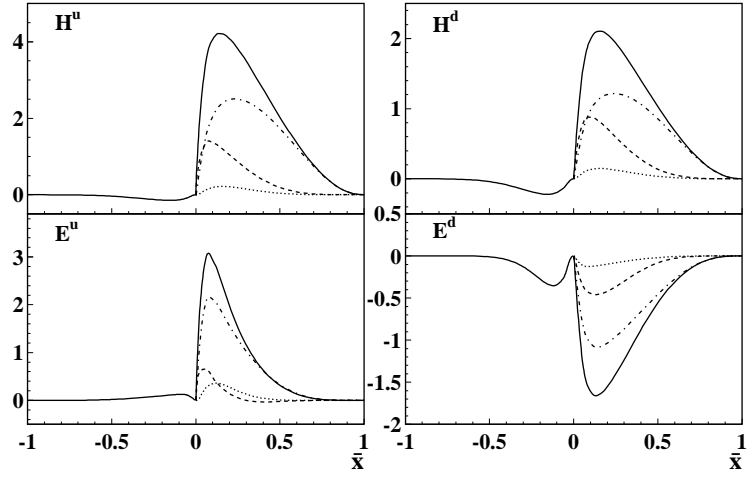,  width=10cm}
\end{center}
\caption{\small The different contributions to the spin-averaged ($H^q$, upper panels) and helicity-flip ($E^q$, lower panels) generalized parton distributions calculated in the meson-cloud model for flavours $u$ (left panels) and $d$ (right panels), at $\xi=0$ and $t=-0.2$ GeV$^2$. Line style as in Fig.~\ref{fig:fig4}.
}
\label{fig:fig6}
\end{figure}

Going beyond the forward limit we first discuss the $t$ dependence of the GPDs at $\xi=0$ by comparing  results in Fig.~\ref{fig:fig4} at $t=0$ with those in Fig.~\ref{fig:fig6} at $t=-0.2$ GeV$^2$. The relative contribution of the different components is not modified by switching on the momentum transfer $t$, only the overall magnitude is decreased. This is in agreement with the common believe that the main part of the $t$ dependence of the GPDs is exhibited by their first moments, i.e. by the quark Dirac and Pauli form factors.

\begin{figure}[ht]
\begin{center}
\epsfig{file=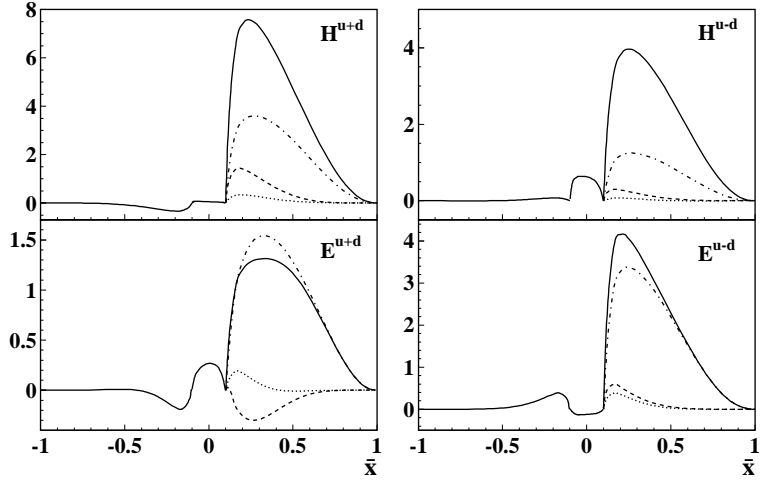,   width=10cm}
\end{center}
\caption{\small Isoscalar ($u+d$, left panels)  and isovector ($u-d$, left panels) combinations of the spin-averaged (upper panels) and helicity-flip (lower panels) generalized parton distributions calculated in the meson-cloud model, at $\xi=0.1$ and $t=-0.2$ GeV$^2$. Line style as in Fig.~\ref{fig:fig4}.
}
\label{fig:fig7}
\end{figure}

\begin{figure}[ht]
\begin{center}
\epsfig{file=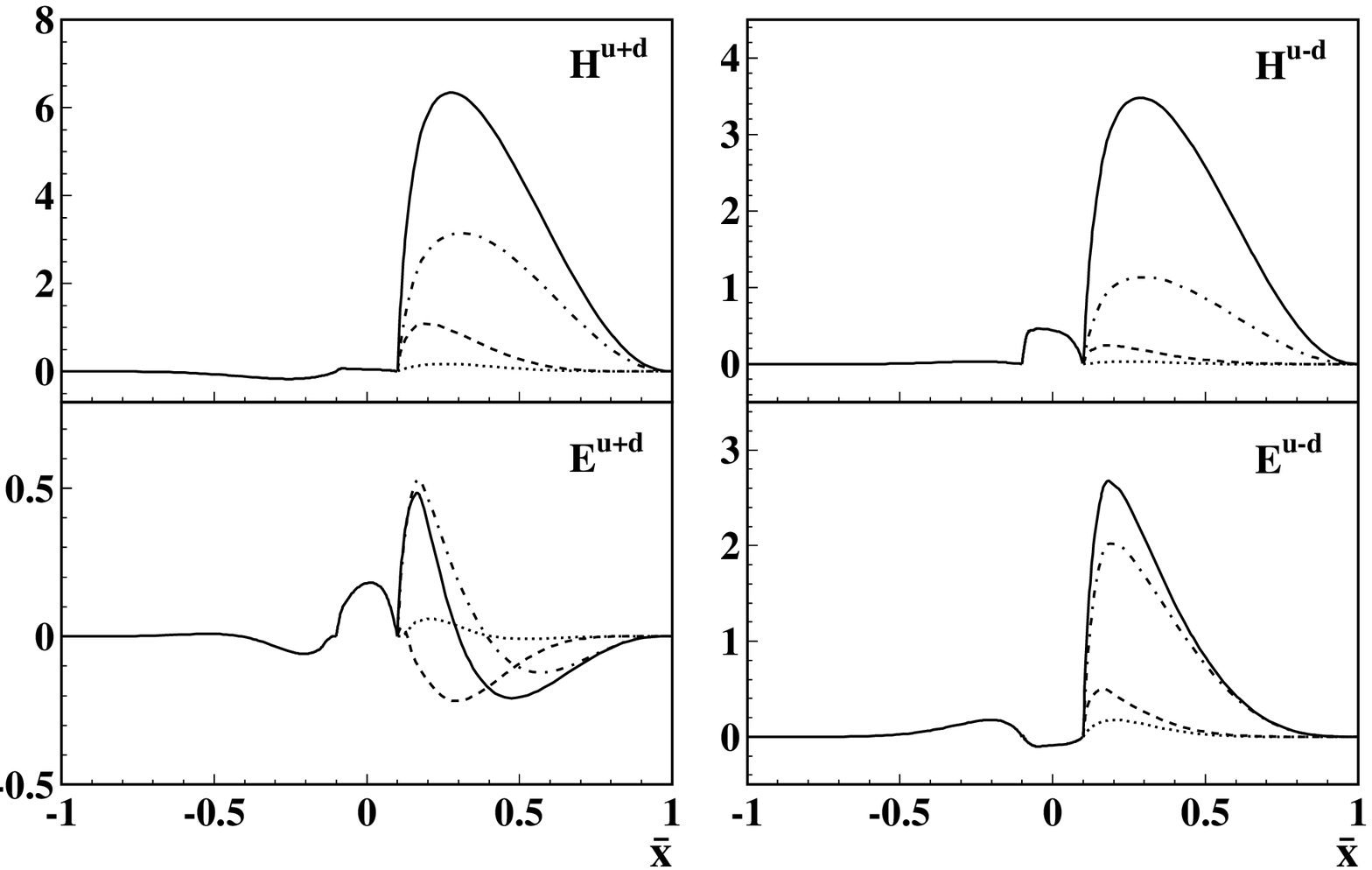,  width=10cm}
\end{center}
\caption{\small Isoscalar ($u+d$, left panels)  and isovector ($u-d$, left panels) combinations of the spin-averaged (upper panels) and helicity-flip (lower panels) generalized parton distributions calculated in the meson-cloud model, at $\xi=0.1$ and $t=-0.5$ GeV$^2$. Line style as in Fig.~\ref{fig:fig4}.
}
\label{fig:fig8}
\end{figure}

\begin{figure}[ht]
\begin{center}
\epsfig{file=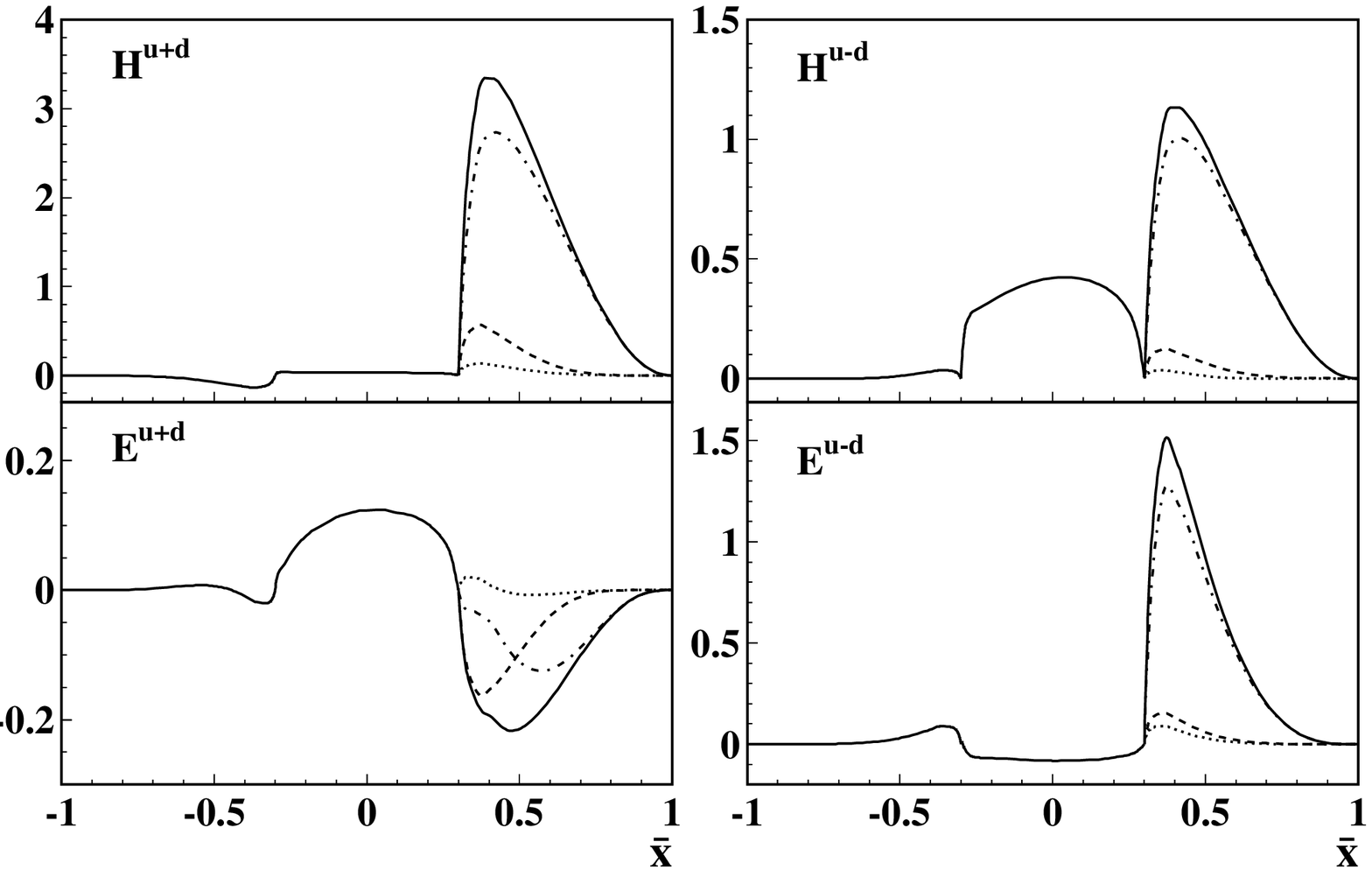,   width=10cm}
\end{center}
\caption{\small Isoscalar ($u+d$, left panels)  and isovector ($u-d$, left panels) combinations of the spin-averaged (upper panels) and helicity-flip (lower panels) generalized parton distributions calculated in the meson-cloud model, at $\xi=0.3$ and $t=-0.5$  GeV$^2$.
}
\label{fig:fig9} 
\end{figure}

With a nonvanishing $\xi$ one can explore the ERBL region with $\vert\overline x\vert\le \xi$ where  in our model only transition amplitudes between the bare-proton and baryon-meson components are contributing. The combined dependence of the isoscalar and isovector combinations of GPDs on $\xi$ and $t$ is shown in Figs.~\ref{fig:fig7}--\ref{fig:fig9}. The $t$ dependence at constant $\xi=0.1$ 
can be extracted from Figs.~\ref{fig:fig7} and \ref{fig:fig8},
 while the $\xi$ dependence at constant $t=-0.5$ GeV$^2$ is deduced 
from Figs.~\ref{fig:fig8} and \ref{fig:fig9}. 

From the isoscalar and isovector combinations of GPDs plotted in Fig.~\ref{fig:fig7} at $\xi=0.1$ and $t=-0.2$ GeV$^2$ we see that GPDs in the ERBL region are rather regular functions over the whole range, with zeros at the endpoints $\overline  x=\pm\xi$. This result is quite different from the oscillatory behaviour predicted by the chiral quark-soliton model~\cite{petrov} where the valence contribution of the discrete level is a smooth function extending into the ERBL region and adding to the sea contribution. Here this is forbidden because the support of the valence contribution is limited to the DGLAP region. In addition, the transition amplitude between the bare-proton and the baryon-meson components vanishes at the boundary of the ERBL region. As discussed in Ref.~\cite{markusthesis}, the points $\overline  x=\pm\xi$ correspond to very peculiar parton configurations involving one parton with vanishing momentum in the initial or final state hadron. As a consequence, when approaching e.g. $\xi$ from below, one probes a quark-antiquark pair with one momentum fraction finite and the other going to zero, a configuration similar to the one of a meson distribution amplitude at its endpoints. On the other hand leading-twist GPDs  must be continuous at $\overline  x=\pm\xi$ to avoid logarithmically divergent scattering amplitudes in DVCS. In fact, this condition is fulfilled here because also approaching $\overline x=\pm\xi$ from the DGLAP region in our model GPDs are constrained to go to zero. This generates a discontinuity of the first derivative of GPDs at $\overline x=\pm\xi$ which, however, is not in contradiction with general principles. Including higher Fock-states of the type suggested in Ref.~\cite{antonuccio} and/or considering evolution to a higher scale will fill up the zero at $\overline x=\pm\xi$.

In the DGLAP region both for $\overline x\ge\xi$ and $\overline x\le -\xi$ no striking difference arises in Fig.~\ref{fig:fig7} for the spin-averaged GPDs $H^{u\pm d}$ with respect to the results in the forward limit shown in Fig.~\ref{fig:fig5}, while for the helicity-flip GPDs the (negative) $d$ contribution coming from the baryon in the $\ket{BM}$ component  is responsible for a broader shape at $\overline x\ge\xi$. The $t$ dependence of the different and opposite contributions to $E^u$ and $E^d$ is also responsible for  the small size and the oscillatory behaviour of $E^{u+d}$ at $t=-0.5$ GeV$^2$ (Fig.~\ref{fig:fig8}). Increasing $\xi$, therefore compressing the support for the valence contribution, this effect is even more visible producing a negative $E^{u+d}$ for $\overline x\ge\xi$ as in Fig.~\ref{fig:fig9}, while the behaviour in the ERBL region remains the same.


\section{Conclusions}
\label{conclusions}

The convolution model for the physical nucleon, where the bare nucleon is dressed by its virtual meson cloud, has a long and successful history in explaining properties such as form factors and parton distributions. In this paper it has been revisited and applied for the first time to study GPDs. A light-front wavefunction overlap representation is obtained in the one-meson approximation by inserting a Fock-state expansion involving a bare nucleon and meson-baryon states. The model fulfills the support condition and general sum rules such as the number, momentum and angular momentum sum rules. Explicit expressions for the unpolarized GPDs have been derived and applied to the case of the meson being a pion and the baryon being either a nucleon or a $\Delta$. 

This meson-cloud model gives the possibility to link GPDs calculated in the light-front formalism to the nucleon description in terms of constituent quarks including a sea contribution already at a low-energy scale and providing a suitable input for the evolution to higher scales. The results presented  in this paper for different kinematics show an important contribution of the meson cloud at low $\overline x$ and a smooth contribution of the sea in the ERBL region and for negative $\overline x$. As an effect of the truncated Fock-state expansion, characteristic nodes occurs at the endpoints of the DGLAP and ERBL region, i.e. at $\overline x=0$ in the forward case and at $\overline x=\pm\xi$ in the off-forward case, where the wavefunction has to vanish. However, this artifact will disappear under evolution of GPDs to higher scales.
In addition, since the contribution in the ERBL region is vanishing in the 
forward limit, it  can not be easily inferred from 
parametrizations
in terms of parton distributions. Therefore, the present calculation
gives new insights to model the off-forward 
features of the GPDs, and can be further used
as input at the hadronic scale to study the behaviour under evolution
at higher scales.

\section*{Acknowledgements}

This research is part of the EU Integrated Infrastructure Initiative Hadronphysics Project under contract number RII3-CT-2004-506078 and was partially supported by the Italian MIUR through the PRIN Theoretical Physics of the Nucleus and the Many-Body Systems. The diagrams in the paper have been drawn using the  Jaxodraw package~\cite{binosi}.


\appendix
\setcounter{section}{0}
\setcounter{equation}{0}
\renewcommand{\theequation}{\Alph{section}.\arabic{equation}}


\section{Baryon scattering amplitude}
\setcounter{equation}{0}
\label{appendixb}

In this section we give the LCWF overlap representation of the baryon 
scattering amplitude given in 
Eq.~(\ref{eq:baryionsa}), and evaluated in the nucleon average frame where the 
plus and transverse components of the momentum of the initial and final baryon 
are given by
\begin{eqnarray}
\label{eq:averagebar}
p^+_B  =
\overline{p}_B^{\, +}\left(1+\frac{\xi}{\overline{y}_B}\right) \eqcm
&\quad&
{\bf p}_{B\perp}=\overline{{\bf p}}_{B\perp}-\frac{{\bf \Delta}}{2}
 \eqcm \nn \\p\,'^+_B  =
\overline{p}\,'^{+}_B\left(1-\frac{\xi}{\overline{y}_B}\right) \eqcm
&\quad&
{\bf p}'_{B\perp}=\overline{{\bf p}}_{B\perp}+\frac{{\bf \Delta}}{2}.
\end{eqnarray}
This scattering amplitude gives a non-vanishing contribution only in the 
region $\xi\leq\overline x\leq 1,$ where it describes the emission of a quark from 
the baryon with momentum fraction $\overline x +\xi/\overline y_B$ of the average plus 
momentum $\overline p_B^+,$ and its reabsorption with $\overline x -\xi/\overline y_B$.
By analogy with the nucleon case, we introduce the names ``baryon-in'' and 
``baryon-out'' for frames
where the incoming and outgoing baryon has zero transverse momentum. 
The  fraction of plus momentum and the transverse components of the momentum
 of partons before the scattering process in the baryon-in frame will be
 denoted with $\tilde \zeta_i$, and $ \tilde{\bold \kappa}_{i\perp}$, 
respectively. The corresponding quantities for the final partons in the 
baryon-out frames are defined by $\hat\zeta'_i$ and 
$\hat{\bold \kappa}_{i\perp}$. Furthermore, defining the average momentum 
of the partons as $\overline k_i=(k^+_1+k'^+_i)/2$, we introduce the fraction 
$\overline\zeta_i=\overline k^+_i/\overline p_B^+=\overline x_i/\overline y_B.$ 
When the active parton is taken out from the baryon it carries a fraction 
$\overline \zeta_j+\xi/\overline y_B$ of the average plus momentum $\overline p_B^+,$ and its 
transverse momentum is $\overline{\bf k}_{j\perp}-{\bf \Delta}_\perp/2$. 
The arguments of the initial-state LCWF in the baryon-in frame are obtained 
from the variables of the partons in the average frame by means of the
 transverse boost in Eq.~(\ref{eq:35})
with parameters 
${\bf b}_\perp={\bf \overline p}_{B\perp}-{\bf \Delta}_\perp/2$ and 
$b^+=\overline p^+_B(1+\xi/\overline y_B)$. 
Furthermore, by using the spectator condition, $k'_i=\overline k_i=k_i$, one obtains
\be
\begin{array} {ll}
\displaystyle \tilde \zeta_i= 
\frac{\overline \zeta_i}{1+{\xi}/{\overline y_B}} \eqcm \qquad& \displaystyle
\tilde{\bold \kappa}_{\perp i}=\overline{\bf k}_{\perp i}
                        -\frac{\overline \zeta_i}{1+{\xi}/{\overline y_B}}\,
\left(\overline{\bf p}_{B\perp}-\frac{{\bf\Delta}_\perp}{2}\right) \eqcm
\qquad \qquad \mbox{for }i\neq j \eqcm \nn\\
\displaystyle\tilde \zeta_j=\frac{\overline \zeta_j
+{\xi}/{\overline y_B}}{1+{\xi}/{\overline y_B}} \eqcm \qquad& \displaystyle
\tilde{\bold \kappa}_{\perp j}=\overline{\bf k}_{\perp j}
                        -\overline{\bf p}_{B\perp}
                        +\frac{1-\overline \zeta_j}{1+{\xi}/{\overline y_B}}\,
      \left(\overline{\bf p}_{B\perp}-\frac{{\bf\Delta}_\perp}{2}\right)
\eqpt
\end{array}
\label{eq:tilde-args-bar}
\ee

In the final state, the active parton has a fraction of the average plus 
momentum $\overline p^+_B$ of the baryon equal to $\overline{\zeta_j}-\xi/\overline y_B$, and 
a transverse momentum $\overline{\bf k}_{j\perp}+{\bf \Delta}_\perp/2$. 
The arguments of the final-state LCWF in the baryon-out frame are obtained 
from the variables of the partons in the average frame by means of the 
transverse boost in Eq.~(\ref{eq:35}) 
with parameters ${\bf b}_\perp={\bf \overline p}_{B\perp}
+{\bf \Delta}_\perp/2$ and $b^+=\overline p^+_B(1-\xi/\overline y_B)$ and are given by
\be
\begin{array}{ll}
\displaystyle \hat \zeta'_i= \frac{\overline \zeta_i}{1-{\xi}/{\overline y_B}} \eqcm 
\qquad& \displaystyle
\hat{\bold{\kappa}}'_{i\perp }=\overline{\bf k}^\adj_{i\perp }
                        -\frac{\overline \zeta_i}{1-{\xi}/{\overline y_B}}\,
                 \left(\overline{\bf p}_B+\frac{{\bf\Delta}_\perp}{2}\right) \eqcm
\qquad \qquad \mbox{for }i\neq j \eqcm \nn\\
\displaystyle\hat \zeta'_j=
\frac{\overline \zeta_j-{\xi}/{\overline y_B}}{1-{\xi}/{\overline y_B}} \eqcm \qquad& 
\displaystyle
\hat{\bold\kappa}'_{j\perp }=\overline{\bf k}^\adj_{j\perp }
                        -\overline{\bf p}_{B\perp}
                        +\frac{1-\overline \zeta_j}{1-{\xi}/{\overline y_B}}\,
          \left(\overline{\bf p}_{B\perp}+\frac{{\bf\Delta}_\perp}{2}\right)
\eqpt
\end{array} 
\label{eq:hat-args-bar}
\ee

Using a similar procedure as in the calculation of the valence-quark 
contribution to the scattering amplitude in the region 
$\xi\leq \overline x\leq 1,$ one finds that the LCWF overlap representation of 
$ F^{q/B}_{\lambda\lambda'}$ is given by
\begin{eqnarray}
F^{q/B}_{\lambda'\lambda}
\left(\frac{\overline x}{\overline y_B},\frac{\xi}{\overline y_B},{\bf \Delta}_\perp\right) 
& = &
\frac{1}{(1-{\xi^2}/{\overline y_B^2})}
\sum_{j=1}^3\,\sum_{\lambda_i\tau_i}\delta_{s_jq}
\int[{\rm d}\overline \zeta]_3
[{\rm d}^2\overline{{\bf k}}_{\perp}]_3\,
\delta\left(\frac{\overline x}{\overline y_B}-\overline \zeta_j\right)
\nonumber\\
& &{}\times 
\Psi^{B,[f]\,*}_{\lambda'}(\{\hat \zeta'_i, \hat{{\bold\kappa}}'_{i\perp};
\lambda_i,\tau_i\}) 
\Psi^{B,[f]}_\lambda(\{\tilde \zeta_i,\tilde{\bold \kappa}_{i\perp};
\lambda_i,\tau_i\}).
\label{eq:dglapbaryon}
\eea


\section{Meson generalized parton distributions}
\label{appendixc}
\setcounter{equation}{0}

In this section we derive the LCWF overlap representation of the meson scattering amplitude in Eq.~(\ref{eq:mesongpd}), with the momenta of the incoming and outgoing meson in the average frame given by
\begin{eqnarray}
\label{eq:averagemes}
p^+_M  =
\overline{p}_M^{\, +}\left(1+\frac{\xi}{\overline{y}_M}\right) \eqcm
&\quad&
{\bf p}_{M\perp}=\overline{{\bf p}}_{M\perp}-\frac{{\bf \Delta}}{2}
 \eqcm \nn \\p^+_M  =
\overline{p}\,'^{+}_M\left(1-\frac{\xi}{\overline{y}_M}\right) \eqcm
&\quad&
{\bf p}'_{M\perp}=\overline{{\bf p}}_{M\perp}+\frac{{\bf \Delta}}{2}.
\end{eqnarray}

In the following we separately discuss the contributions in the DGLAP region for positive and negative $\overline x.$ 

\vspace{0.5cm}

\noindent {\it {Region $\xi\leq\overline x\leq 1$ }}

\vspace{0.5cm}

The derivation of $F^{q/M}$ in terms of meson LCWFs  goes along the same lines as in the case of the baryon discussed in App.~\ref{appendixb}. The final result is given by
\begin{eqnarray}
F^{q/M}_{\lambda'\lambda}
\left(\frac{\overline x}{\overline y_M},\frac{\xi}{\overline y_M},{\bf \Delta}_\perp\right) 
& = &
\frac{1}{\sqrt{1-{\xi^2}/{\overline y_M^2}}}
\sum_{j=1}^2\,\sum_{\lambda_i\tau_i}\delta_{s_jq}
\int[{\rm d}\overline \zeta]_2\,[{\rm d}^2\overline{{\bf k}}_{\perp}]_2\,
\delta\left(\frac{\overline x}{\overline y_M}-\overline \zeta_j\right)
\nonumber\\
& &{}\times 
\Psi^{M,[f]\,*}_{\lambda'}(\{\hat \zeta'_i,\hat{{\bold\kappa}}'_{i\perp};
\lambda_i,\tau_i\}) 
\Psi^{M,[f]}_\lambda(\{\tilde \zeta_i,\tilde{\bold\kappa}_{i\perp};
\lambda_i,\tau_i\}) ,
\label{eq:dglapmes2}
\eea
where $\{\tilde \zeta_i\},\, \{\tilde{\bold\kappa}_{i\perp}\}$ are the coordinates of the initial partons in the meson-in frame. They are related to the coordinates in the average frame by the same transformation as in Eq.~(\ref{eq:tilde-args-bar}), with the substitution
 $\overline y_B\rightarrow \overline y_M$, and ${\bf p}_{B\perp}\rightarrow {\bf p}_{M\perp}$.  Likewise, in Eq.~(\ref{eq:dglapmes2}) $\{\hat \zeta'_i\},\, \{\hat{\bold\kappa}'_{i\perp}\}$ are the coordinates of the final partons in the meson-out frame and satisfy the same relations as in Eq.~(\ref{eq:hat-args-bar}) with  $\overline y_B$ and ${\bf p}_{B\perp}$ replaced by $\overline y_M$ and 
 ${\bf p}_{M\perp}$, respectively.
\vspace{0.5cm}

\noindent {\it {Region $-1\leq\overline x\leq -\xi$ }}

\vspace{0.5cm}

In this region, the meson contribution to the scattering amplitude is given by 
\begin{eqnarray}
F^{q/M}_{\lambda'\lambda}
\left(\frac{\overline x}{\overline y_M},\frac{\xi}{\overline y_M},{\bf \Delta}_\perp\right) 
& = &
-
\frac{1}{\sqrt{1-{\xi^2}/{\overline y_M^2}}}
\sum_{j=1}^2\,\sum_{\lambda_i\tau_i}\delta_{s_j\bar q}
\int[{\rm d}\overline \zeta]_2\,[{\rm d}^2\overline{{\bf k}}_{\perp}]_2\,
\delta\left(\frac{\overline x}{\overline y_M}+\overline \zeta_j\right)
\nonumber\\
& &{}\times 
\Psi^{M,[f]\,*}_{\lambda'}(\{\hat \zeta'_i, \hat{{\bold\kappa}}'_{i\perp};
\lambda_i,\tau_i\}) 
\Psi^{M,[f]}_\lambda(\{\tilde \zeta_i,\tilde{\bold\kappa}_{i\perp};
\lambda_i,\tau_i\})
\label{eq:dglapmes3}
\eea
where the arguments of the LCWFs are given by the same expression as in the region $\xi\le \overline x\le 1$.

\section{Vertex functions}
\label{appendixa}

In this Appendix we work out the case of the 
$N\rightarrow N\pi$ and $N\rightarrow \Delta\pi$ transitions
 in the light-front formalism. 
Vertex functions for such transitions can be found in several 
places~(see, e.g., Refs.\cite{HSS96, DHSS97,Speth98}), but it is convenient to 
show their derivation.

The light-front vectors are defined as
\begin{equation}
A^\mu=(A^-, A^+,{\bf A}_\perp),
\end{equation}
with
\begin{equation}
A^\pm=A^0\pm A^3,\quad {\bf A}_\perp=(A^1,A^2).
\end{equation}
We also use the notations $A_{R,L}=A^1\pm iA^2$ and $\tilde A=(A^+,{\bf A}_\perp)$.

The light-front nucleon spinors $u_\lambda(\tilde p)$ are 
given by
\begin{equation}
u_{1/2}(\tilde p)=\frac{1}{\sqrt{2p^+}}
\left(
\begin{array}{c}
p^+ + m\\
p_R\\
p^+-m\\
p_R
\end{array}
\right),
\qquad
u_{-1/2}(\tilde p)=\frac{1}{\sqrt{2p^+}}
\left(
\begin{array}{c}
-p_L\\
p^+ + m\\
p_L\\
m-p^+
\end{array}
\right).
\label{eq:spinor}
\end{equation}

The gamma matrices are defined as in Ref.~\cite{Bjo65}.

A similar expansion for the $\Delta$ field involves the Rarita-Schwinger 
spinors given by
\begin{eqnarray}
u^\mu_{3/2}(\tilde p)&=&\epsilon^\mu_{+1}(\tilde p)\,u_{1/2}(\tilde p),\nn\\
u^\mu_{1/2}(\tilde p)&=&\sqrt{\frac{2}{3}}\epsilon^\mu_{0}(\tilde p)\,u_{1/2}
(\tilde p)
+\sqrt{\frac{1}{3}}\epsilon^\mu_{+1}(\tilde p)\,u_{-1/2}(\tilde p)
,\nn\\
u^\mu_{-1/2}(\tilde p)&=&\sqrt{\frac{2}{3}}\epsilon^\mu_{0}(\tilde p)\,u_{-1/2}
(\tilde p)
+\sqrt{\frac{1}{3}}\epsilon^\mu_{-1}(\tilde p)\,u_{1/2}(\tilde p)
,\nn\\
u^\mu_{-3/2}(\tilde p)&=&\epsilon^\mu_{-1}(\tilde p)\,u_{-1/2}(\tilde p),
\end{eqnarray}
where the polarization vectors are given by
\begin{eqnarray}
\epsilon^\mu_{+1}(\tilde p)&=&\left(-\sqrt{2}\frac{p_R}{p^+},0,
(-\frac{1}{\sqrt{2}},
-\frac{i}{\sqrt{2}})\right),\nn\\
\epsilon^\mu_{0}(\tilde p)&=&\frac{1}{m}\left(\frac{{\bf p}_\perp^2-m^2}{p^+},
p^+,
{\bf p}_\perp\right),\nn\\
\epsilon^\mu_{-1}(\tilde p)&=&
\left(\sqrt{2}\frac{p_L}{p^+},0,(\frac{1}{\sqrt{2}},
-\frac{i}{\sqrt{2}})\right).
\end{eqnarray}

In the evaluation of the transition amplitudes we need the following 
expressions of scalar products
\begin{eqnarray}
(\epsilon^\mu_{+1})^*(\tilde p_\Delta)(p_N - p_\Delta)_\mu&=&
-\frac{k_L}{\sqrt{2}y},\nn\\
(\epsilon^\mu_{0})^*(\tilde p_\Delta)(p_N - p_\Delta)_\mu&=&
\frac{1}{2 M_\Delta y}\left({\bf k}_\perp^2-M^2_\Delta+y^2M_N^2\right),\nn\\
(\epsilon^\mu_{-1})^*(\tilde p_\Delta)(p_N - p_\Delta)_\mu&=&
\frac{k_R}{\sqrt{2}y},
\end{eqnarray}
where the relation ${\bf p}_{\Delta\perp}={\bf k}_\perp+y{\bf p}_{N\perp}$ has been used.

The final results for the transition amplitudes $N\rightarrow B\pi$ with 
nucleon helicity $\lambda=\oneh$ 
are given in Table I.
The corresponding results for the vertex functions 
with nucleon helicity $\lambda=-\oneh$ are given by
\begin{eqnarray}
V_{\lambda',0}^{\lambda\,(N,N\pi)}(y,{\bf k}_\perp) &=&(-1)^{1/2-\lambda'}\,
 V_{-\lambda'}^{-\lambda\,(N,N\pi)}(y,\hat{{\bf k}}_\perp), \nonumber\\
V_{\lambda'}^{\lambda\,(N,\Delta\pi)}(y,{\bf k}_\perp) 
&=&(-1)^{3/2-\lambda'}\,
 V_{-\lambda'}^{-\lambda\,(N,\Delta\pi)}(y,\hat{{\bf k}}_\perp), 
\end{eqnarray}
where $\hat{{\bf k}}_\perp = (k_x,-k_y)$.

Our calculation for the vertex functions is in agreement with the results 
of Ref.~\cite{DHSS97}. 
However, we found discrepancies with the results reported in
Ref.~\cite{Speth98}. These differences in the vertex functions  do not affect 
the splitting functions of Eqs.~(\ref{eq:57}) and (\ref{eq:59}) which enter
in the convolution formulas of parton distributions, 
but are important for the transverse-momentum dependence
of the probability amplitudes $\phi_{\lambda\lambda'}$ which enter in the
convolution formulas for the GPDs.

\begin{table}
\begin{center}
\begin{tabular}{||c|c|c||}
\hline\hline
$\lambda\rightarrow\lambda'$& $V(N,N\pi)$& $V(N,\Delta\pi)$
\\
\hline
& &  \\
${1\over 2}\rightarrow{3\over 2}$
&&
$-ig_{ N\Delta \pi}\frac{k_L}{\sqrt{2y}y}(M_\Delta+yM_N)$\\
& &\\
${1\over 2}\rightarrow{1\over 2}$&
$ig_{ NN\pi }\frac{M_N(1-y)}{\sqrt{y}}$&
$ig_{ N\Delta \pi}\frac{1}{\sqrt{6y}yM_\Delta}
[{\bf k}_\perp^2(2M_\Delta+yM_N)+(M_\Delta+yM_N)^2(yM_N-M_\Delta)]$\\
& &\\
${1\over 2}\rightarrow-{1\over 2}$&
$-ig_{ NN\pi }\frac{k_R}{\sqrt{y}}$&
$-ig_{ N\Delta \pi}\frac{k_R}{\sqrt{6y}yM_\Delta}[{\bf k}_\perp^2+(M_\Delta
+yM_N)
(yM_N-2M_\Delta)]$\\
& &\\
${1\over 2}\rightarrow-{3\over 2}$
&&
$-ig_{ N\Delta\pi}\frac{k_Rk_R}{\sqrt{2y}y}$
  \\
& & \\
\hline\hline
\end{tabular}
\end{center}
\caption{ 
Vertex functions for $N\rightarrow N\pi$ and $N\rightarrow \pi\Delta$.
\label{table1} }
\end{table}



\end{document}